\documentclass[aps,floats]{revtex4}
\usepackage{amsmath,amssymb}
\usepackage{graphicx,epsfig}

\begin{document}
\bibliographystyle {plain}

\def\oppropto{\mathop{\propto}} 
\def\opsimeq{\mathop{\simeq}}
\def\opoverderline{\mathop{\overline}}
\def\operarrow{\mathop{\longrightarrow}}
\def\opsim{\mathop{\sim}}

\def\fig#1#2{\includegraphics[height=#1]{#2}}
\def\figx#1#2{\includegraphics[width=#1]{#2}}


\title{ On the multifractal statistics of the local order parameter 
at random critical points : \\
application to wetting transitions with disorder } 


\author{ C\'ecile Monthus and Thomas Garel }
 \affiliation{Service de Physique Th\'{e}orique, CEA/DSM/SPhT\\
Unit\'e de recherche associ\'ee au CNRS\\
91191 Gif-sur-Yvette cedex, France}

\begin{abstract}

Disordered systems present multifractal properties at criticality.
In particular, as discovered by Ludwig 
(A.W.W. Ludwig, Nucl. Phys. B 330, 639 (1990))
on the case of diluted two-dimensional Potts model,
the moments $\overline{\rho^q(r)}$ of the local order parameter $\rho(r)$
scale with a set $x(q)$ of non-trivial exponents $x(q) \neq q x(1)$.
In this paper, we revisit these ideas to incorporate
 more recent findings: 
(i) whenever a multifractal measure $w(r)$ normalized over space
$ \sum_r w(r)=1$ occurs in a random system,
it is crucial to distinguish between the typical values
and the disorder averaged values of
 the generalized moments $Y_q =\sum_r w^q(r)$, since they may
scale with different generalized dimensions $D(q)$ and $\tilde D(q)$
(ii) as discovered by Wiseman and Domany (S. Wiseman and E. Domany,
Phys Rev E {\bf 52}, 3469 (1995)), the presence of an infinite
correlation length induces a lack of self-averaging
at critical points for thermodynamic observables, in particular 
for the order parameter.
After this general discussion valid for any random critical point,
 we apply these ideas to
  random polymer models that can be studied numerically
for large sizes and good statistics over the samples.
We study the bidimensional wetting or the Poland-Scheraga DNA model
with loop exponent $c=1.5$ (marginal disorder) and $c=1.75$
(relevant disorder).
Finally, we argue that the presence of finite Griffiths ordered clusters at 
criticality determines the asymptotic value 
$x(q \to \infty) =d$ and the minimal value $
 \alpha_{min}=D(q \to \infty)=d-x(1) $ of the typical multifractal
spectrum $f(\alpha)$.

\end{abstract}

\maketitle

\section{Introduction}

Among the various areas where multifractality occurs 
(see for instance \cite{halsey,Pal_Vul,Stan_Mea,Aha,Meakin,
harte,duplantier_houches} and references therein), 
the case of critical points in the presence of frozen
disorder is of particular interest.
The idea that multifractality occurs at criticality
has been first established for quantum Anderson localization transitions
\cite{Weg,Cas_Pel} and has been the subject of 
very detailed studies \cite{Jan,Huck,Mirlin}.
For the directed polymer in a random medium
in dimension $1+3$, where a disorder-induced localization/delocalization 
occurs, the multifractal properties studied recently 
\cite{DPmultif} are very similar to the case of Anderson transitions.
In the field of spin models, 
the most studied case seems to be the 
two-dimensional diluted $q$-state Potts model,
where multifractality was discovered by Ludwig \cite{Ludwig}
via conformal field theory using perturbation theory in
the parameter $(q-2)$ governing the disorder relevance.
This work has motivated numerical studies for various values
of $q$ \cite{Jac_Car,Ols_You,Cha_Ber,PCBI}.
The idea of multifractality has been also
proposed in other disordered models like spin-glasses and random
field spin systems \cite{Sourlas,Thi_Hil,Par_Sou},
and has been studied numerically for spin glasses on
diamond hierarchical lattice \cite{brazil}.
Finally, for disordered quantum spin-chains,
it turns out that 
the statistics of critical correlation functions is described by
``multiscaling'',
which is even stronger than multifractality \cite{multiscaling}.
This is because these disordered quantum spin-chains
 are actually governed by ``Infinite disorder fixed points''
\cite{revueigloi}.

So the presence of multifractality
at criticality seems generic in disordered systems.
However, following \cite{Ludwig}, most studies on classical
disordered models have
focused on the statistics of two-point correlation functions,
whereas multifractality already occurs at the level of one-point
functions like the order parameter or the energy density \cite{Ludwig}.
In particular,
the moments of the local order parameter scale
with a set $x(n)$ of non-trivial exponents $x(n) \neq n x(1)$.
In this paper, we revisit these ideas in the light of
 more recent findings, concerning the possible
differences between the exponents for typical and averaged values,
and the lack of self-averaging
of thermodynamic observables at criticality \cite{domany95,AH,domany}.
We then study the multifractal statistics of the order parameter
in random polymer models that can be studied numerically
for large sizes and good statistics over the samples.

The paper is organized as follows.
In Section \ref{theory}, we summarize the outcome of previous works
concerning the multifractal statistics and the lack of self-averaging
at random critical points. In the remainder of the paper,
we apply these ideas to wetting transitions with disorder.
These random polymer models are presented in Section
 \ref{models}. Our numerical results
on the statistics of the local order parameter 
are given respectively in Sections \ref{resc1.75} for loop exponent $c=1.75$
(relevant disorder) and in Section \ref{resc1.5} for loop exponent $c=1.5$
(marginal disorder). In Section 
 \ref{boundary}, we discuss the influence of boundary conditions
on the multifractal spectrum. 
Section \ref{conclusion} contains our conclusions.

\section{ Multifractal statistics of the local order parameter 
at criticality}

\label{theory}

\subsection{ Disorder averaged moments of local order parameter }

Let $\rho(r;i,L)$ be the local order parameter at site $r$,
in a finite disordered sample $i$ of volume $L^d$ in dimension $d$.
This local order parameter is usually defined 
in terms of a thermal average, for instance 
$\rho(r;i,L)= < \sigma(r;i,L)> $ in disordered ferromagnets,
$\rho(r;i,L)= < \sigma(r;i,L)>^2 $ in spin-glasses.
For the random polymer models described in Section \ref{models},
 $\rho(r;i,L)$ corresponds to the contact density of monomer $r$
(Eq. \ref{rhowetting}).
Let $\rho(i,L)$ denote 
the spatial average over all points $r$ of the sample
\begin{eqnarray}
\rho(i,L) \equiv \frac{1}{L^d} \sum_{r \in L^d} 
 \rho(r;i,L)
\label{rhospatialav}
\end{eqnarray}

In a pure system, the exponent $x_{pure}$ that governs the decay of
the spatial average $\rho_{pure}(L)$ also describes the decay
of the local parameter $\rho_{pure}(r;L)$ for any point $r$ in the bulk
\begin{eqnarray}
 \rho_{pure}(L)
  \sim \frac{1}{L^{x_{pure}}}  \sim \rho_{pure}(r \in bulk;L)
\label{pure}
\end{eqnarray}

In a disordered sample however, the spatial heterogeneity of the disorder
induces a spatial heterogeneity in the local order parameter
 $\rho(r;i,L)$ at criticality.
In particular, there exists 
a family of non-trivial exponents $x(q) \ne q x(1)$ \cite{Ludwig}
for the disorder averaged powers of the local order parameter
\begin{eqnarray}
\overline{ [ \rho(r;i,L) ]^q }
 \oppropto_{L \to \infty} \frac{1}{L^{x(q)}}  
\label{rhoqav}
\end{eqnarray}
In this formula, it is convenient to consider $q$ as a continuous
real parameter to probe also non-integer moments and even negative moments.

\subsection{ Introduction of a normalized multifractal measure   }

Since the multifractal formalism is usually defined
for a normalized probability measure \cite{halsey},
it is convenient to construct a probability measure
from the non-normalized observables one is interested in
 \cite{Dup_Lud,Pook,Jan,Huck}.
Here for the local order parameter, one defines 
in each sample $(i)$ the following spatial weights 
\begin{eqnarray}
w(r;i,L) 
= \frac{ \rho(r;i,L ) }{ \displaystyle \sum_{r' \in L^d} 
\rho(r';i,L ) } 
= \frac{ \rho(r;i,L ) }{ L^d \ \rho(i,L) } 
\label{defw}
\end{eqnarray}
normalized to
\begin{eqnarray}
\sum_{r \in L^d} w(r;i,L) =1
\label{normaw}
\end{eqnarray}
So $w(r;i,L)$ represents the contribution of the site $r$ to
the order parameter of the sample $i$ of size $L^d$.
The statistics of these weights can be studied via
the following generalized moments 
\begin{eqnarray}
Y_q(i,L) = \sum_{r \in L^d} \big[ w(r;i,L) \big]^q 
\label{defyq}
\end{eqnarray}
Deep in the ordered phase, where the order parameter $\rho(i,L)$ 
is finite as $L \to \infty$, the weights are expected to be
of the same order $1/L^d$ (see Eq. \ref{defw}).
The decay of the generalized moments then follow the simple scaling
\begin{eqnarray}
Y_q(i,L) \vert_{ordered phase} \sim \frac{1}{L^{(q-1)d}}
\label{yqordered}
\end{eqnarray}
At criticality however, the generalized moments $Y_q$
will display multifractality, with a priori different exponents
for typical and averaged values.

\subsection{ Typical generalized dimensions $D(q)$   }

At criticality, the decay of typical values
define a series of generalized exponents $\tau(q)=(q-1) D(q) $
\begin{equation}
Y_q^{typ}(L) \equiv e^{ \overline{ \ln Y_q(i,L)} }   \sim 
\frac{1}{ L^{  \tau(q) }} = \frac{1}{ L^{ (q-1) D (q) } }
\label{tctyp}
\end{equation}
The exponents $D(q)$ represent generalized dimensions \cite{halsey} :
$D(0)$ represent the dimension of the support of the measure,
here it is simply given by the space dimension
\begin{equation}
D(0)=d
\label{dzero}
\end{equation}
 $D(1)$ is usually called the information dimension \cite{halsey} ,
since it describes the behavior of 
the 'information' entropy
\begin{equation}
s(i,L) \equiv - \sum_{ r \in L^d } w(r;i,L) \ln  w(r;i,L)
= - \partial_q Y_q(i,L) \vert_{q=1} \simeq  D(1) \ln L
\label{entropy}
\end{equation}
Finally $D(2)$ is called the correlation dimension \cite{halsey}
and describes the decay of
\begin{equation}
Y_2^{typ}(L) \equiv e^{ \overline{ \ln Y_2(i,L)} }  \simeq 
 L^{ - D (2) } 
\label{y2d2}
\end{equation}

\subsection{ Typical singularity spectrum $f(\alpha)$   }

In the multifractal formalism, the singularity spectrum $f(\alpha)$
is given by the Legendre transform of $\tau(q)$ \cite{halsey}
via the standard formula
\begin{eqnarray}
 q && =f'(\alpha) \\
 \tau(q) && = \alpha q - f(\alpha)
\label{legendre}
\end{eqnarray}
The physical meaning of $f(\alpha)$ is that the number ${\cal N}_L(\alpha)$
of points $r$ where the weight $w(r;i,L)$
scales as $L^{-\alpha}$ typically behaves as 
\begin{eqnarray}
{\cal N}_L^{typ}(\alpha) \equiv e^{ \overline{ \ln {\cal N}_L(\alpha) }  }
 \propto L^{f(\alpha)}
\label{nlalpha}
\end{eqnarray}
So the Legendre transform of Eq. (\ref{legendre}) corresponds to
the saddle-point calculus in $\alpha$ of the following expression
\begin{equation}
Y_q^{typ}(L) \sim \int d\alpha \ L^{f(\alpha)} \ L^{- q \alpha} 
\label{saddle}
\end{equation}
The general properties of 
the singularity spectrum $f(\alpha)$ are as follows \cite{halsey} :
it is positive $f(\alpha) \geq 0$
 on an interval $[\alpha_{min},\alpha_{max}]$
where $\alpha_{min}=D(q=+\infty)$ is the minimal singularity exponent
and $\alpha_{max}=D(q=-\infty)$ is the maximal singularity exponent.
It is concave $f''(\alpha)<0$. 
It has a single maximum at some value $\alpha_0$ where
$f(\alpha_0)=D(q=0)$, so here (Eq. \ref{dzero})
\begin{equation}
f(\alpha_0) =D(q=0)=d
\end{equation}
The singularity exponent $\alpha_0$ is thus the typical value
\begin{equation}
\alpha_0 =\alpha_{typ}
\label{alphatyp}
\end{equation}

However, the singularity that yields the leading contribution
to the normalization 
$Y_1 = \sum_r w(r)=1$ of the measure is the singularity exponent
given by the information dimension $D(1)$ of Eq. \ref{entropy}
\begin{equation}
\alpha_1=f(\alpha_1)=D(1)
\label{alpha1}
\end{equation}

\subsection{ Generalized dimensions $ \tilde D(q)$ 
defined from disorder averaged values   }

Following \cite{halsey}, many authors consider that
the singularity spectrum has a meaning only for $f(\alpha) \geq 0$
\cite{Jan,Huck}. However, when
multifractality arises in random systems, disorder-averaged values
may involve other generalized exponents 
\cite{mandelbrot,Chh_neg,Jen_neg,has_dup}
than the typical values (see Eq. \ref{tctyp}).
In quantum localization transitions, these exponents were 
denoted by  $\tilde \tau(q)=
(q-1) \tilde D(q) $ in \cite{Mirlin}
and we will follow these notations
\begin{equation}
\overline{ Y_q(i,L)}  \simeq
\frac{1}{ L^{ \tilde \tau(q) }} = \frac{1}{ L^{ (q-1) \tilde D (q) } }
\label{tcav}
\end{equation}
For these disorder averaged values, the corresponding singularity
spectrum $\tilde f(\alpha)$ defined by 
\begin{eqnarray}
 \overline{ {\cal N}_L(\alpha) }  
 \propto L^{{\tilde f}(\alpha)}
\label{nlalphaav}
\end{eqnarray}
may become negative $\tilde f(\alpha)<0$ 
\cite{mandelbrot,Chh_neg,Jen_neg,has_dup,Mirlin}
to describe rare events.

\subsection{ Wiseman-Domany lack of self-averaging at criticality }

To make the link between the exponents $x(q)$ defined from the
powers of the local order parameter (Eq. \ref{rhoqav})
and the multifractal exponents of the normalized weights (Eq. \ref{defw})
\cite{Dup_Lud}, one needs to
 use the equivalence between spatial average
and disorder averages for the local order parameter.

\subsubsection{ Lack of self-averaging of extensive thermodynamic observables}

 In disordered systems off-criticality, 
the densities of extensive thermodynamic 
observables are self-averaging,
because  the finiteness of the correlation length $\xi(T)$
allows to divide a large sample into independent large
sub-samples.
At criticality however, this 'subdivision' argument breaks down
because of the divergence of $\xi(T_c)=\infty$ at $T_c$, and a  
 lack of self-averaging has been found at criticality
 whenever disorder is relevant
\cite{domany95,AH,domany}.
More precisely, for a given observable $X$,
it is convenient to define its normalized width as
\begin{equation}
\label{defratiodomany}
R_X(T,L) \equiv \frac{ \overline { X_i^2(T,L)} - ( \overline{X_i(T,L)})^2
}{ ( \overline{X_i(T,L)})^2 } 
\end{equation}
To be more specific, in ferromagnets, the observable $X$
can be the magnetization $M$, the susceptibility $\chi$,
the singular parts of the energy or of the specific heat \cite{domany}.
In terms of the correlation length  $\xi(T)$, the following behaviour of
$R_X(T,L)$ is expected \cite{AH,domany}  :

(i) off criticality, the correlation length $\xi(T)$
is finite. For $L \gg \xi(T)$, 
the system can be then divided into nearly independent sub-samples
and this leads to `Strong Self-Averaging' 
\begin{equation}
R_X(T,L) \sim \frac{1}{ L^d} \ \ \hbox{ off  criticality  for 
 $L \gg \xi(T)$ } 
\label{strongsa}
\end{equation}

(ii) in the critical region, when $L \ll \xi(T)$, 
the system cannot be divided anymore into nearly independent sub-samples.
In particular at $T_c$ where $\xi(T_c)=\infty$,
one can have either `weak self-averaging' for  irrelevant
disorder according to the Harris criterion \cite{harris}, i.e.
whenever the pure specific heat exponent $\alpha_{pure}=2-d \nu_{pure}$
is negative
\begin{equation}
R_X(T_c(\infty),L)
 \sim L^{ \frac{\alpha_{pure}}{\nu_{pure}}}  \ \ \hbox{ for  irrelevant
disorder ($\alpha_{pure} <0$)  } 
\label{weaksa}
\end{equation}
or `No Self-Averaging'
\begin{equation}
R_X(T_c(\infty),L) \sim Cst \ \ \hbox{ for   random  critical points  }
\label{nosa}
\end{equation}
Note that for the marginal case $\alpha_{pure}=0$
 from the point of view of the Harris criterion,
 the power governing the `weak self-averaging' of Eq. \ref{weaksa}
vanishes, so the ratio $R_X(T_c(\infty),L)$ can either remain finite
as in Eq. \ref{nosa} or vanish logarithmically.

 \subsubsection{ Application to the powers of the local order parameter }

Let us now apply these results to the spatial averages of powers of the 
local order parameter 
\begin{eqnarray}
\rho_q(i,L) \equiv \frac{1}{L^d} \sum_{r \in L^d} 
\left[ \rho(r;i,L) \right]^q
\label{rhospatialavq}
\end{eqnarray}
that generalizes Eq. \ref{rhospatialav} to arbitrary $q$.

In the ordered phase, the `Strong Self-Averaging' property of Eq.
\ref{strongsa} means
\begin{eqnarray}
\rho_q(i,L) \vert_{T<T_c} \opsimeq_{L \to \infty}
 r_q(T) + \frac{v_q(i)}{L^{d/2}}
\label{ordered}
\end{eqnarray}
where the leading term $r_q(T)$ is non-random and coincides
with the disorder-averaged value in the thermodynamic limit $L \to \infty$
\begin{eqnarray}
 r_q(T) = \lim_{L \to \infty} 
\left( \overline{ \left[ \rho(r;i,L) \right]^q} \right) 
\label{orderedleading}
\end{eqnarray}
and where $v_q(i)$ is a random variable depending on the sample $(i)$.

At criticality, the `No Self-Averaging' result of Eq. \ref{nosa}
means that the spatial averages defined in Eq. \ref{rhospatialavq}
behave asymptotically as
\begin{eqnarray}
 \rho_q(i,L) 
   \opsimeq_{L \to \infty} \frac{u_q(i)}{L^{x(q)}}
\label{rhoaviq}
\end{eqnarray}
where the exponent $x(q)$ 
is the exponent governing the decay of the disorder-averaged 
q-moment $\overline{ \rho^q(r;i,L )}$ of Eq. \ref{rhoqav}.
and where $u_q(i)$ is a random variable of order $O(1)$
depending on the sample $(i)$.

In the following, we will use these result to understand the relations
between the exponents for non-normalized observables
and for the normalized measure. 
It will be useful to introduce
the rescaled variable
\begin{eqnarray}
 u_q(i,L) \equiv L^{x(q)}  \rho_q(i,L)
\label{defui}
\end{eqnarray}
that remains a random variable $u_q(i)$ of order $O(1)$ 
in the limit $L \to \infty$.

\subsection{ Relation between the exponents $x(q)$ and $\tau(q)$ }

In terms of the local order parameter $\rho(r;i,L )$,
the generalized moments $Y_q(i,L)$ reads from Eqs \ref{defw}
and \ref{defyq}
\begin{eqnarray}
Y_q(i,L) = \frac{ \displaystyle  \sum_{r \in L^d} \rho^q(r;i,L ) }
{ \left( \displaystyle  \sum_{r' \in L^d} \rho(r';i,L ) \right)^q}  
\label{yqil}
\end{eqnarray}

From Eq \ref{defui} concerning the spatial averages of Eq.
\ref{rhospatialavq}, one obtains
\begin{eqnarray}
Y_q(i,L)   = 
\frac{ L^{d-x(q)} u_q(i,L)}{  \left( L^{d-x(1)} u_1(i,L) \right)^q} 
= L^{d-x(q)-q(d-x(1))} \ \frac{u_q(i,L)}{ \left( u_1(i,L) \right)^q}
\label{yqalpha}
\end{eqnarray}
The typical values of the random variables $u_q(i,L)$
in the limit $L \to \infty$ 
are of order $O(1)$ and thus the exponents $\tau(q)$
governing the typical values of Eq. \ref{tctyp} read
\begin{equation}
 \tau(q)  \equiv (q-1)  D (q)  = x(q)-d +q (d-x(1))
\label{lien1}
\end{equation}
or equivalently the typical generalized dimensions read
\begin{equation}
 D (q)  = d - \frac{ q x(1) -x(q) }{q-1}  
\label{lien2}
\end{equation}
So these relations written in Ref \cite{Dup_Lud}
 relate the exponents $x(q)$
of disorder-averaged moments of the local order parameter 
(Eq. \ref{rhoqav} )
to the typical exponents $\tau(q)$ of the normalized measure 
(Eq. \ref{tctyp}). 
However, the exponents $\tilde \tau(q)$ of Eq. \ref{tcav}
cannot be simply
related to $(x(q),\tau(q))$, since the disorder average of
 Eq. \ref{yqalpha} may involve a $L$-dependent rare event contribution
of the random variables 
$\overline{\frac{u_q(i,L)}{ \left( u_1(i,L) \right)^q} }$,
in particular for large $q$, since $q$ enters as a power
 in the denominator.

\subsection{ Conclusion for the statistics of the 
local order parameter }

Let us now come back to our starting point,
namely the local order parameter $\rho(r;i,L)$
at site $r$. Using Eq. \ref{defui} for $q=1$,
we obtain in terms of the weight of Eq. \ref{defw}
\begin{eqnarray}
\rho(r;i,L ) = w(r;i,L) \left[  L^d \ \rho(i,L)  \right] 
=   w(r;i,L) L^{d-x(1)} u_1(i,L)
\label{rholocal}
\end{eqnarray}
where $u_1(i,L)$ is a random variable of order $O(1)$.
So the interpretation of the singularity spectrum $f(\alpha)$
for the weights given in Eq. \ref{nlalpha}
can be rephrased as follows :
${\cal N}_L(\alpha) \propto L^{f(\alpha)}$ represents the
the number
of points $r$ where the weight $w(r;i,L)$
scales as $L^{-\alpha}$, i.e.
the number
of points $r$ where the local order parameter scales as
$\rho(r;i,L ) \sim L^{-y}$ with 
\begin{eqnarray}
y= \alpha-d+x(1)
\end{eqnarray}
In particular, the typical exponent $y_{typ}$
governing the logarithmic average
\begin{eqnarray}
\overline{ \ln \rho(r;i,L)  }
 \oppropto_{L \to \infty} - y_{typ} \ln L
\label{rhotyp}
\end{eqnarray}
is related to the typical value $\alpha_{typ}$ of Eq. \ref{alphatyp} by
\begin{eqnarray}
y_{typ}= \alpha_{typ}-d+x(1)
\end{eqnarray}
Similarly, the minimal $y_{min}$
and maximal $y_{max}$ exponents are
related to $\alpha_{min}=D(q=+\infty)$ and
 $\alpha_{max}=D(q=-\infty)$. In particular, since
the minimum value $y_{min}$ cannot be negative, one has
the bound
\begin{eqnarray}
y_{min}= \alpha_{min}-d+x(1) \geq 0
\label{ymin}
\end{eqnarray}

\subsection{ Critical region  }

Both in quantum localization \cite{Jan,Huck,Mirlin}
and in disordered ferromagnets \cite{Ludwig}, the multifractal statistics
exactly at $T_c$ is expected to coexist with a single correlation
length exponent $\nu$ outside $T_c$.
More precisely, the powers of the local order parameters 
are expected to follow the finite-size scaling form
in the critical region around $T_c$ (Eq. \ref{rhoqav} )
\begin{eqnarray}
\overline{ [ \rho(r;i,L;T) ]^q }
 \oppropto_{L \to \infty} \frac{1}{L^{x(q)}}  \  \Phi_q \left( (T-T_c)
L^{1/\nu} \right)
\label{rhoqavfss}
\end{eqnarray}
For $T<T_c$, the convergence to finite-values 
$\overline{[\rho(r;L=\infty;T) ]^q}$  in the $L \to \infty$ limit yields 
\begin{equation}
\overline{ [ \rho(r;L=\infty;T) ]^q }
 = (T_c-T)^{ \tilde \beta(q) } \ \ \ {\rm with } \
\ \tilde \beta(q)=\nu \tilde \tau (q)
\end{equation}
So the presence of a multifractal spectrum $x(q) \ne q x(1)$ at criticality
corresponds to non trivial exponents $\beta(q) \ne q \beta(1)$
for the powers of the local order parameters in
the ordered phase.

\section{ Reminder on Wetting and Poland-Scheraga transitions }

 \label{models}

\subsection{Wetting and Poland-Scheraga models  }

Wetting transitions are in some sense the simplest phase transitions,
since they involve linear systems \cite{mfisher}. Let us consider  a
one-dimensional random walk (RW) of $2L$ steps, starting at $z(0)=0$,
with increments $z(r+1)-z(r)=\pm 1$. The random walk
is constrained to remain  in the upper half plane $z \geq 0$, but
gains an adsorption energy $\epsilon_{r}$ if $z(r)=0$. More
precisely, the model is defined by the partition function 
\begin{equation}
Z_{wetting}(2L) = \displaystyle \sum_{RW} 
\exp \left( \beta \displaystyle \sum_{1 \leq r \leq L}
 \epsilon_{r}\delta_{z_{2 r},0}  \right) 
 \label{zwetting}
  \end{equation}
with inverse temperature $\beta =1/T$.
In the pure case $\epsilon_{r}=\epsilon_0$, there exists a 
continuous phase transition between
a localized phase at low temperature, characterized by an extensive
number of contacts at $z=0$, and a delocalized phase at high
temperature. 

The Poland-Scheraga (PS) model of DNA denaturation \cite{Pol_Scher} is
closely related to the wetting model. It describes 
the configuration of the two complementary chains
as a sequence of bound segments and open loops.
Each loop of length $l$ has a polymeric entropic weight ${\cal N} (l)
\sim \mu^l /l^c $, whereas each contact at position $r$ has a
Boltzmann weight $e^{- \beta \epsilon_{r}}$. We assume that the
two chains are bound at $r=1$ and $r=L$. The partial
partition function $Z_{PS}(r)$ with bound ends then satisfies the
simple recursion relation 
\begin{equation}
Z_{PS}(r)=  e^{-\beta \epsilon_{r} }  
  \sum_{r'=1}^{r-1}   {\cal N}(r-r') Z_{PS}(r')
\label{recursion}
\end{equation}

The wetting model (\ref{zwetting}) corresponds to a Poland-Scheraga
model with loop exponent $c=3/2$ (this exponent comes from the first
return distribution of a one-dimensional random walk).  For DNA
denaturation, the 
appropriate value of the loop exponent $c$ has been the source of some
debate. Gaussian loops in $d=3$ dimensions are
characterized by $c=d/2=3/2$. The role of self avoidance
within a loop was taken into account by Fisher \cite{Fisher}, and
yields the bigger value $c=d \nu_{SAW} \sim 1.76$, where $\nu_{SAW}$
is the SAW radius of gyration exponent in $d=3$. More recently, 
Monte Carlo simulations of self avoiding walks \cite{Barbara1,Carlon}
 and theoretical arguments \cite{Ka_Mu_Pe} pointed 
towards a value $c>2$.

\subsection{ Disorder relevance as a function of the loop exponent $c$}

The Harris criterion concerning the stability
of pure second order transitions with respect to disorder
relies on  the sign of the specific heat exponent 
\begin{equation}
\alpha_P=2-\nu_P = \frac{2c-3}{c-1}
\label{harriscriterion}
\end{equation}
Disorder is thus irrelevant for $1<c<\frac{3}{2}$, marginal for $c=3/2$
 relevant for $\frac{3}{2}<c<2$. 
Poland-Scheraga models are thus particularly interesting
to study disorder effects on pure phase transitions, since the parameter $c$
allows to study, within a single model,
 the various cases of second order transition with respectively
marginal/relevant disorder according to the Harris criterion,
or first-order transition.
From this point of view, it is reminiscent of the 2D Potts model,
where the pure critical properties vary with the parameter $q$ :
the transition is second order for $q<4$, the Ising case $q=2$
corresponding to the marginal case of the Harris criterion, whereas
the transition becomes first order for $q>4$.
 The marginal case $c=\frac{3}{2}$ has been studied for a long time
\cite{FLNO,Der_Hak_Van,Bhat_Muk,Ka_La,Cu_Hwa,Ta_Cha,wetting2005,PS2005}
and is of special interest since it corresponds to two-dimensional wetting
as explained above.

\subsection{ Numerical details }

In the following, we will study the multifractal properties
of the local contact density 
\begin{eqnarray}
\rho(r;i,L) \equiv  < \delta_{z_r,0} >_{i,L}
\label{rhowetting}
\end{eqnarray}
representing the probability that the monomer $r$ of the sample $(i)$
of length $L$ is on the interface $z=0$ at criticality $T=T_c$.
We have chosen the same disorder distribution and parameters as in our
previous work \cite{PS2005}, and we have used the same Fixman-Freire
scheme to speed up calculations, as explained in details in
\cite{wetting2005,PS2005}. The results presented below have been
obtained for the following sizes $L$ and the corresponding number
$n_S(L)$ of disordered samples 
\begin{eqnarray}
\frac{L}{10^3} && = 16,32,64,128,256,512  \\
\frac{n_s(L)}{10^4} && = 500, 250, 120, 60,  30, 15
\end{eqnarray}

\section{Multifractal analysis of the wetting transition 
with loop exponent $c=1.75$ }

 \label{resc1.75}

In this Section, we describe our results for the wetting transition
with loop exponent $c=1.75$ that corresponds to relevant disorder
as explained above
(Eq. \ref{harriscriterion}).

\subsection{Exponents $x(q)$ and generalized dimensions $D(q)$ 
and $ \tilde D(q)$ }

\begin{figure}[htbp]
\includegraphics[height=6cm]{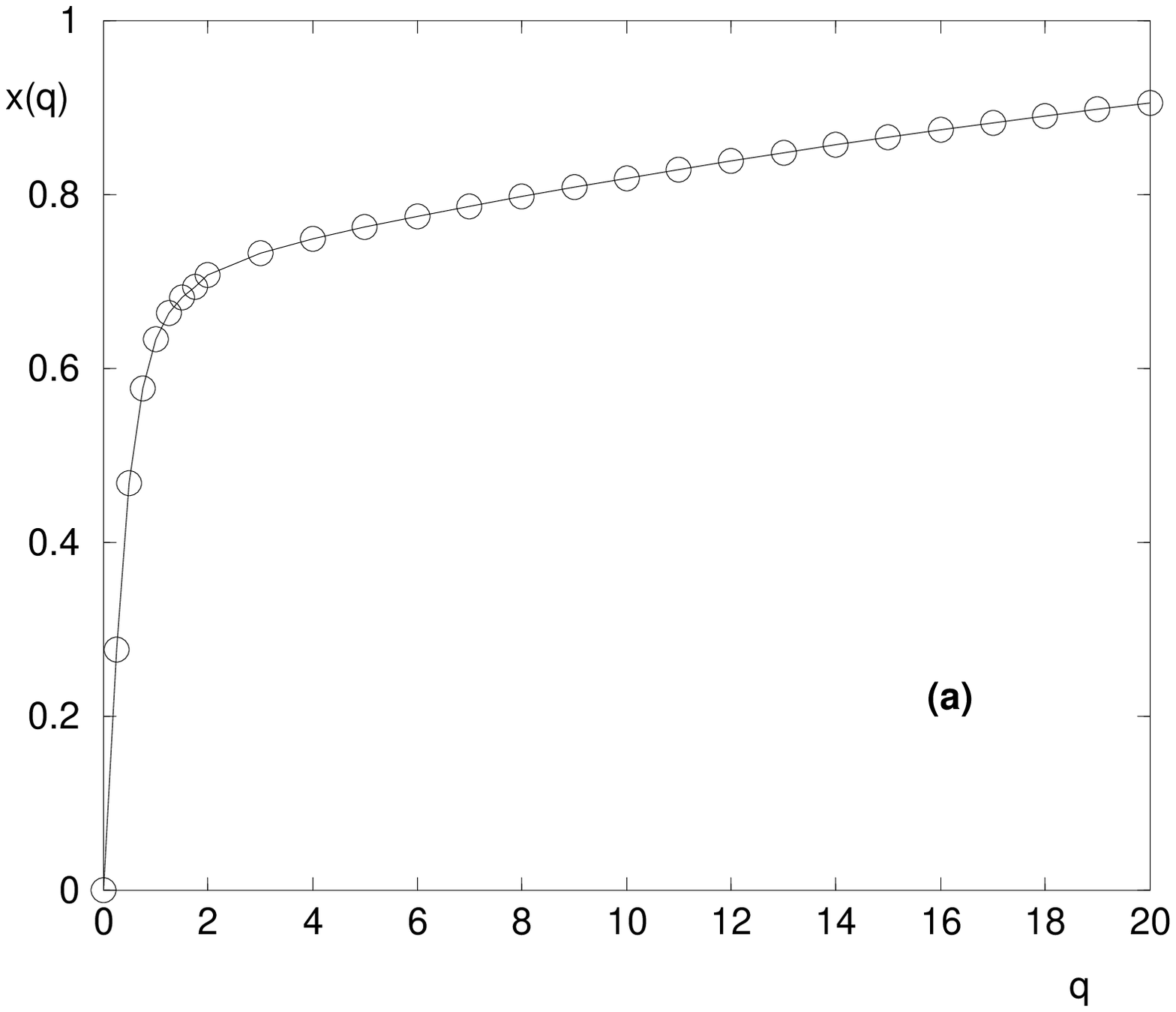}
\hspace{1cm}
\includegraphics[height=6cm]{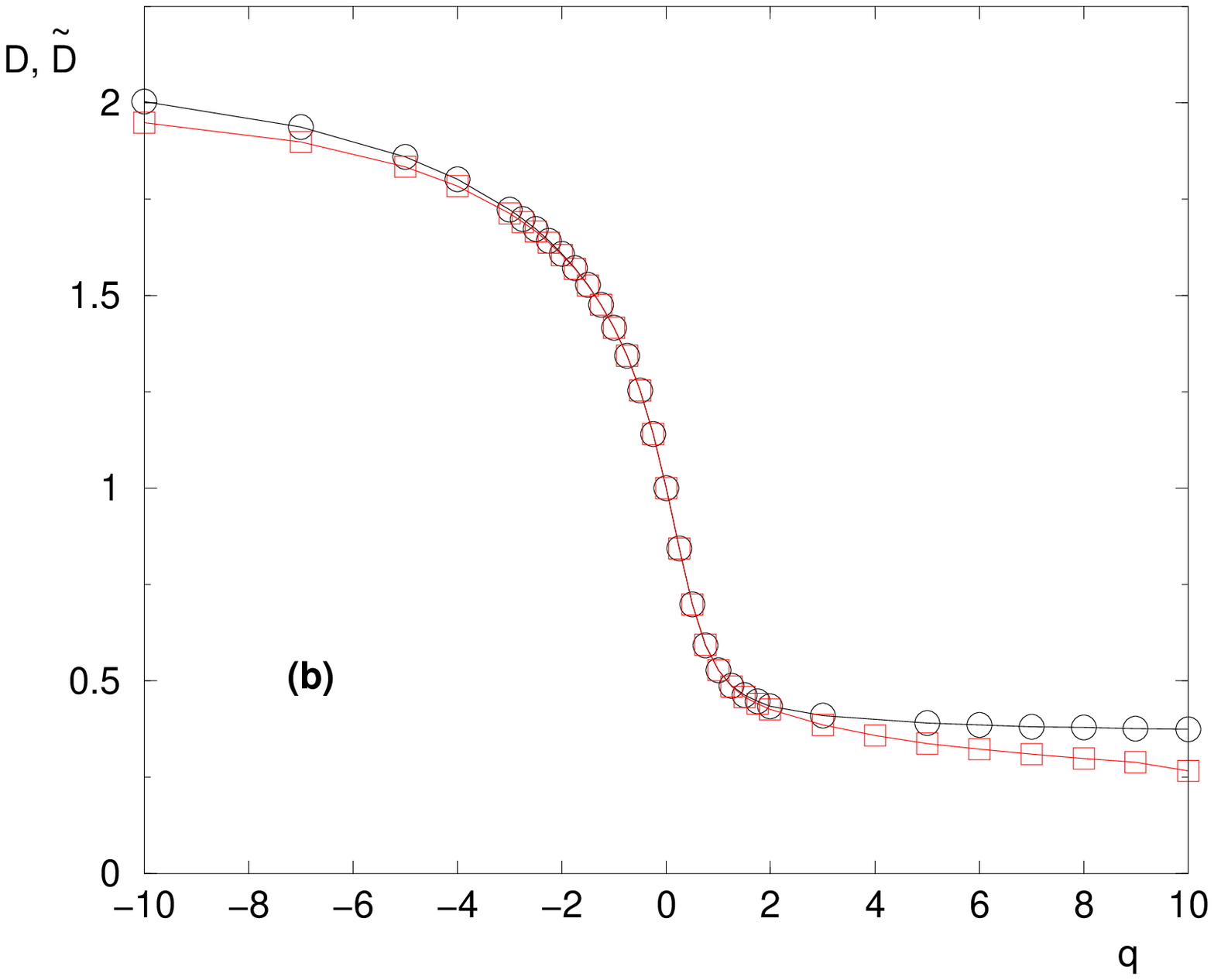}
\caption{(Color online) Wetting with loop exponent $c=1.75$
(a) Exponents $x(q)$ governing the decay of the disorder averaged $q$-th powers of the
local order parameter (Eq. \ref{defxqwett})
(b) Generalized dimensions $D(q)$ ($\bigcirc$) and $\tilde D(q)$ ($\square$) associated respectively
to the typical values (Eq. \ref{tctyp}) and 
to disorder-averaged values (Eq. \ref{tcav}) of the generalized moments
(Eq. \ref{yqilwett}). }
\label{figexpc1.75}
\end{figure}

We show on Fig. \ref{figexpc1.75} (a) 
the scaling dimensions $x(q)$ governing
the disorder-averaged moments of the local contact density
(Eq. \ref{rhoqav})
\begin{eqnarray}
 \overline{ < \delta_{z_r,0} >^q } \propto \frac{1}{ L^{x(q)} }
\label{defxqwett}
\end{eqnarray}
It is strongly non linear $x(q) \ne q x(1)$.
Some particular values are
\begin{eqnarray}
x(q=1) && \simeq 0.63 \\
x(q=2) && \simeq 0.71
\end{eqnarray}
In the large $q$ limit, it saturates towards 
\begin{eqnarray}
x(q \to +\infty) \to 1 
\label{xqinfty1}
\end{eqnarray}
This point will be discussed in details in Section \ref{boundary}.

On Fig. \ref{figexpc1.75} (b), we show the generalized dimensions
$D(q)$ and $\tilde D(q)$ associated respectively
to the typical values (Eq. \ref{tctyp}) and 
to disorder-averaged values (Eq. \ref{tcav}) of the generalized moments
(Eq. \ref{defyq})
\begin{eqnarray}
Y_q(i,L) = \frac{ \displaystyle  \sum_{r} < \delta_{z_r,0} >^q }
{ \left( \displaystyle  \sum_{r' } < \delta_{z_r',0} >  \right)^q}  
\label{yqilwett}
\end{eqnarray}
In particular, the information dimension of Eq. \ref{entropy}
is
\begin{eqnarray}
D(q=1) = {\tilde D}(1) \simeq  0.59
\end{eqnarray}
and the correlation dimension 
of Eq. \ref{y2d2} is
\begin{eqnarray}
D(q=2) \sim {\tilde D}(2) \simeq 0.44
\end{eqnarray}
For $q$ large enough, the two exponents do not coincide anymore
$D(q) \ne {\tilde D}(q)$, as expected from the discussion on Eqs 
\ref{yqalpha}, \ref{lien1}, \ref{lien2}.

\subsection{Typical singularity spectrum $f(\alpha)$ }

\begin{figure}[htbp]
\includegraphics[height=6cm]{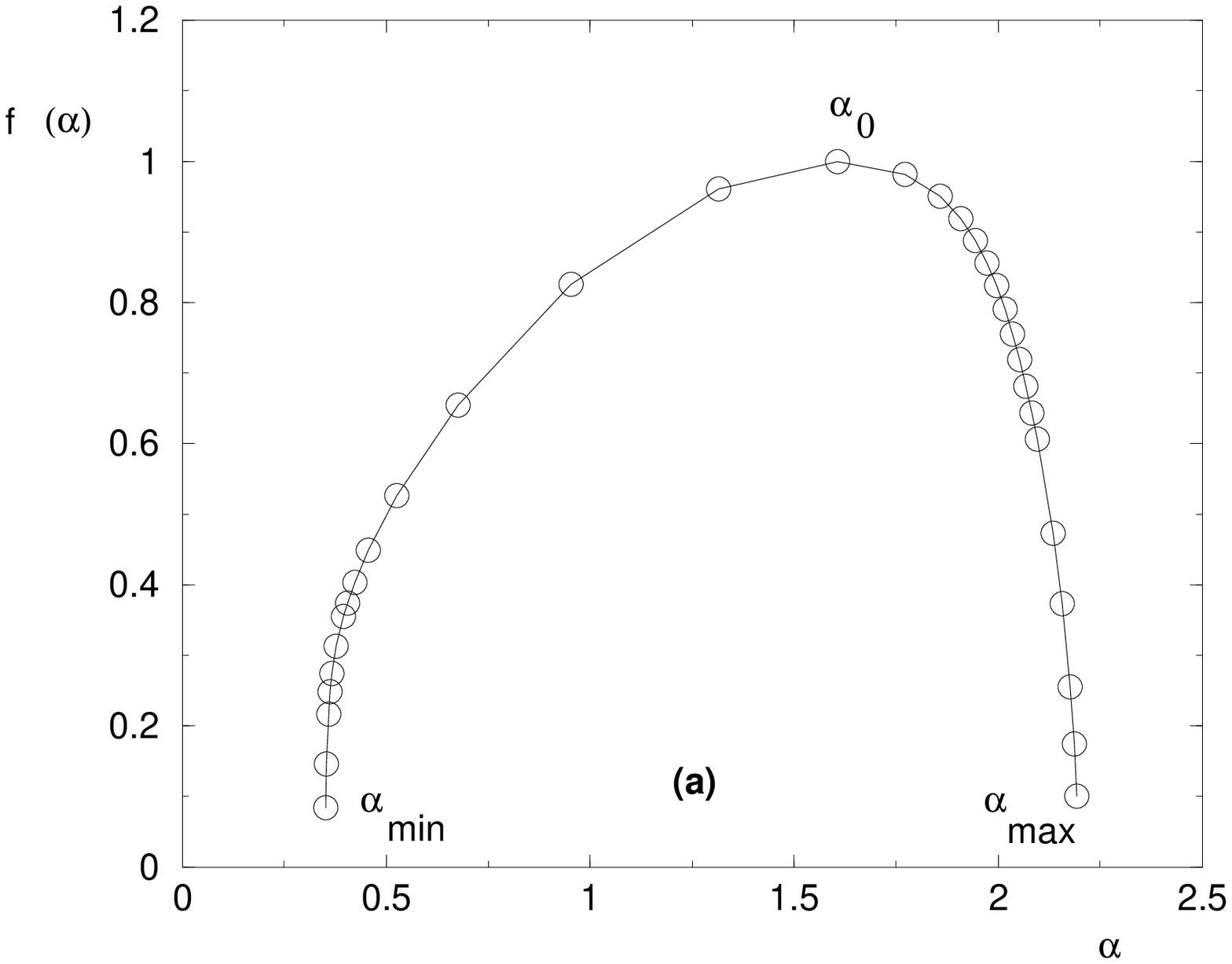}
\hspace{1cm}
\includegraphics[height=6cm]{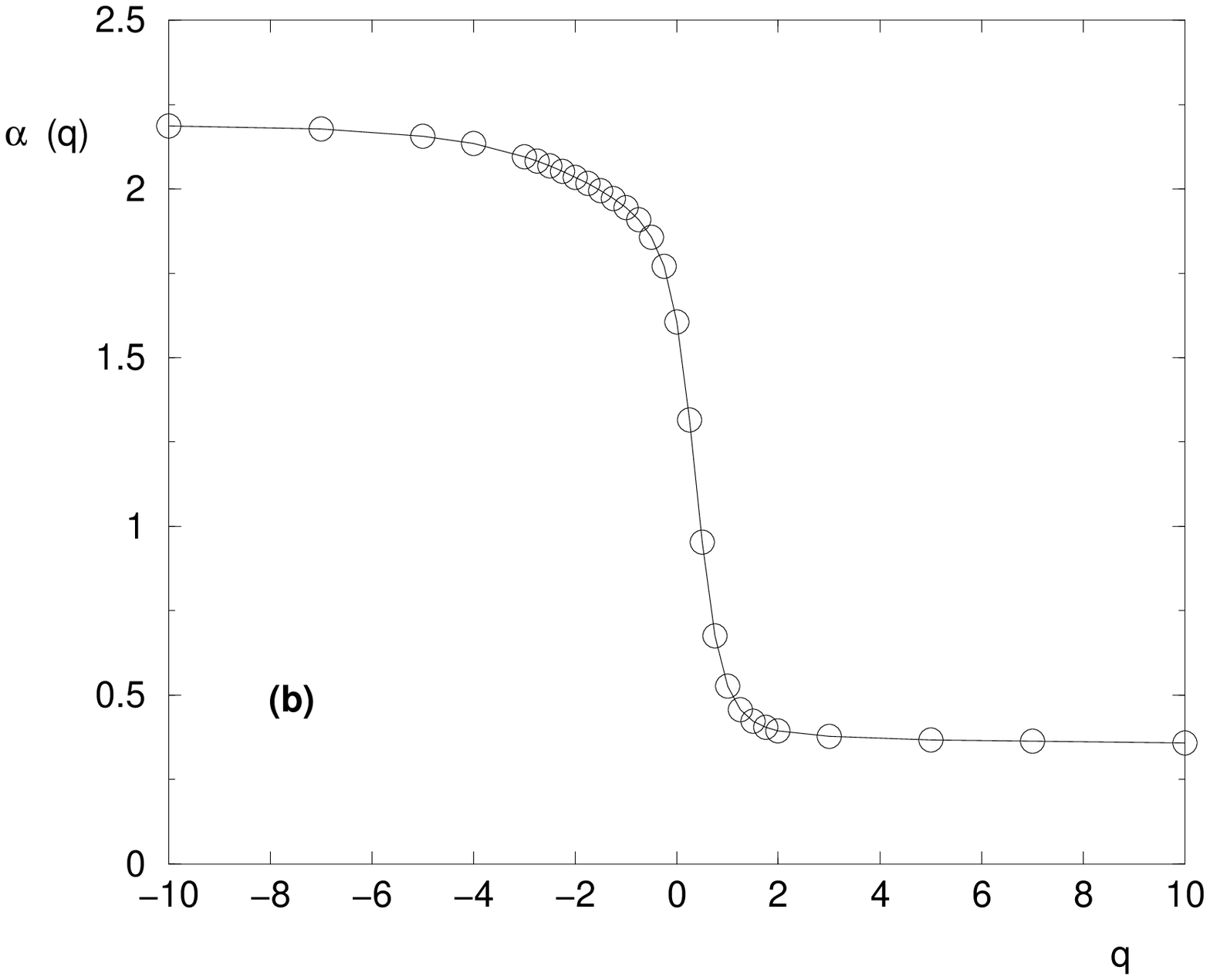}
\caption{(Color online) Wetting with loop exponent $c=1.75$
(a) Typical singularity spectrum $f(\alpha)$ (Eq. \ref{nlalpha}) :
the maximum occurs at $\alpha_0 \sim 1.53 $ which represents the
typical value.
The minimal value is around $\alpha_{min} \simeq 0.36$.
(b) Corresponding curve $\alpha(q)$.  }
\label{figfalphac1.75}
\end{figure}

To measure the typical singularity spectrum introduced in Eq. \ref{nlalpha},
we have used the standard method based on $q$-measures of Ref. \cite{Chh}.
We show on Fig. \ref{figfalphac1.75} (a) 
the curve $f(\alpha)$.
The maximum corresponds to the typical exponent $\alpha_0$ (Eq.
\ref{alphatyp})
\begin{equation}
\alpha_{typ}=\alpha_0 = 1.53
\label{alphatypc1.75}
\end{equation}
The curve is tangent to the diagonal $\alpha=f(\alpha)$ at the point (Eq. \ref{alpha1})
\begin{equation}
\alpha_1=f(\alpha_1)= D(1) \simeq  0.59
\label{alpha1c1.75}
\end{equation}
The minimal value corresponds to
\begin{equation}
\alpha_{min}= D(q=+\infty) \simeq 0.36
\end{equation}
and the maximal value
\begin{equation}
\alpha_{max}= D(q=-\infty) \simeq 2.22
\end{equation}

On Fig. \ref{figfalphac1.75} (b), we show the corresponding curve $\alpha(q)$
representing the dominant exponent $\alpha$ that contribute
to the $q$-generalized moment (Eq. \ref{saddle}). 

\subsection{ Disorder-averaged singularity spectrum ${\tilde f}(\alpha)$  }

\begin{figure}[htbp]
\includegraphics[height=6cm]{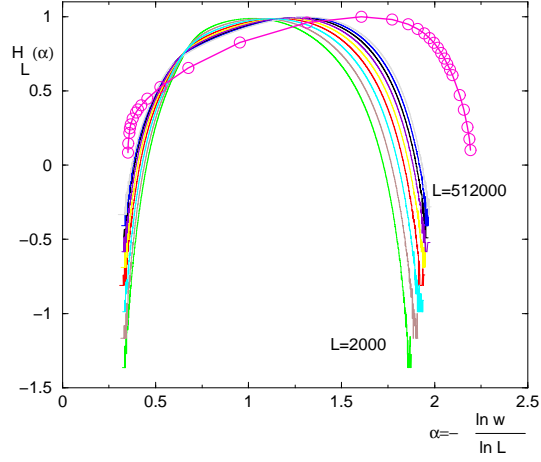}
\hspace{1cm}
\caption{(Color online) Wetting with loop exponent $c=1.75$ :
the histograms $H_L(\alpha)$ of the rescaled variable $\alpha = -
\frac{ \ln w(r;i,L) }{ \ln L}$
shown for $2.10^3 \leq L \leq 512.10^3$
converge extremely slowly (as $1/\ln(L)$) towards the typical singularity spectrum
$f(\alpha)$ ($\bigcirc$) measured with the method of Ref. \cite{Chh}.
 }
\label{figftildec1.75}
\end{figure}

In contrast to the method of \cite{Chh} that allows
to measure numerically the typical spectrum, we are not aware
of an efficient method to measure the 
 disorder-averaged singularity spectrum ${\tilde f}(\alpha)$.
We have thus measured the probability distributions $H_L(\alpha) $
of the rescaled weights
\begin{equation}
\alpha = - \frac{ \ln w(r;i,L) }{ \ln L}
\end{equation}
in analogy with similar numerical measures of the multifractal spectrum
from the statistics of correlation function in disordered
Potts models \cite{Ols_You,Cha_Ber}.
Our results presented on Fig. \ref{figftildec1.75} 
show that the convergence towards the typical spectrum $f(\alpha)$
in the positive region $\alpha>0$ is extremely slow.
In particular, the convergence of the most probable exponent $\alpha_{mp}(L)$
towards the typical value $\alpha_0$ is extremely slow, of order $1/\ln(L)$.

\subsection{ Histograms of the `information' entropy and of $Y_2$ }

\begin{figure}[htbp]
\includegraphics[height=6cm]{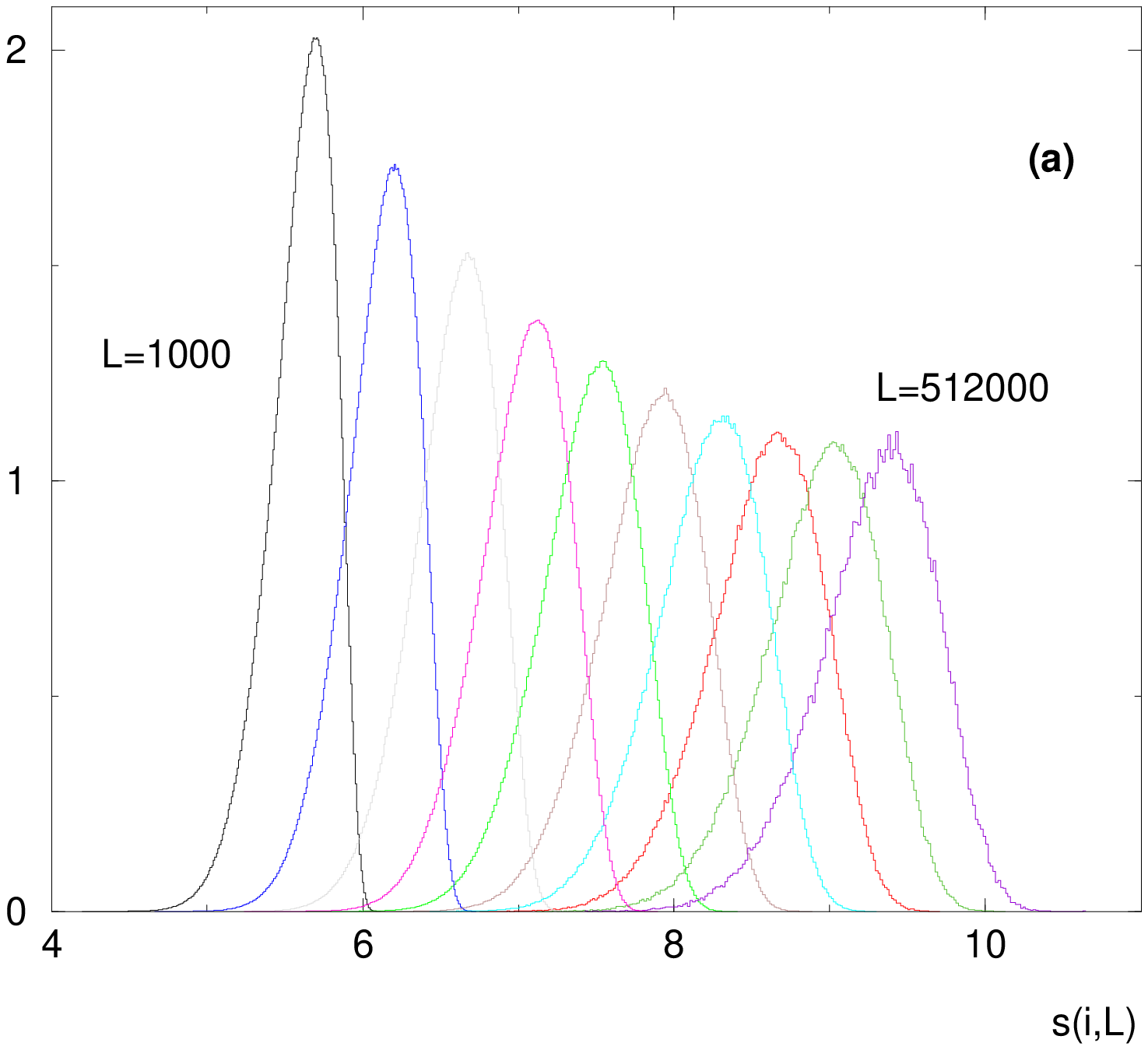}
\hspace{1cm}
\includegraphics[height=6cm]{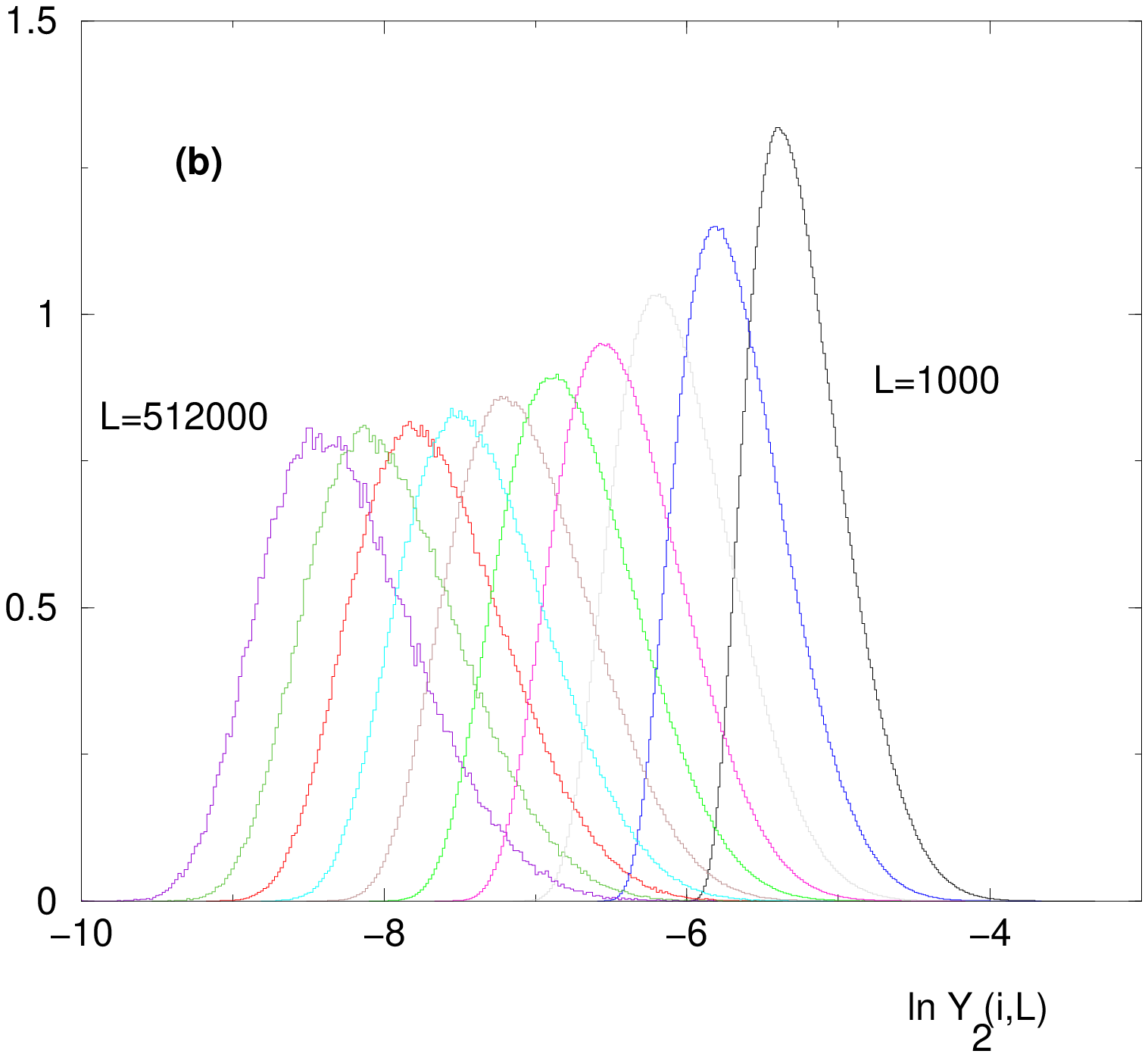}
\caption{(Color online) Wetting with loop exponent $c=1.75$
(a) histogram over the samples of the information entropy
$s(i,L)$ defined in Eq. \ref{entropy} for sizes $10^3 \leq L \leq
512.10^3$ :  the averaged value grows as $\overline{s(i,L) }
\simeq  D(1) \ln L$
(b) histogram over the samples of $\ln Y_2(i,L)$ (Eq.  \ref{defyq} )
 for sizes $10^3 \leq L \leq
512.10^3$ :  the averaged value behaves as $\overline{  \ln Y_2(i,L) } \simeq  - D(2) \ln L$.
}
\label{fighistoc1.75}
\end{figure}

We show on Fig. \ref{fighistoc1.75} (a) the histogram over the samples $(i)$
of the 'information' entropy $s(i,L)$ defined in Eq. \ref{entropy} :
as $L$ grows, the averaged value grows logarithmically (Eq.  \ref{entropy})
\begin{equation}
\overline{s(i,L) } \simeq  D(1) \ln L
\label{entropyav}
\end{equation}
whereas the width converges towards a constant value.

Similarly, we show on Fig. \ref{fighistoc1.75} (b) 
the histogram over the samples $(i)$
of $\ln Y_2(i,L)$ (Eq.  \ref{defyq} ) :
as $L$ grows, the averaged value grows logarithmically (Eq. \ref{y2d2})
\begin{equation}
\overline{  \ln Y_2(i,L) } \simeq  - D(2) \ln L
\label{lny2av}
\end{equation}
whereas the width converges towards a constant value.

\subsection{Statistics of the spatial averaged order parameter
 over the samples }

\begin{figure}[htbp]
\includegraphics[height=6cm]{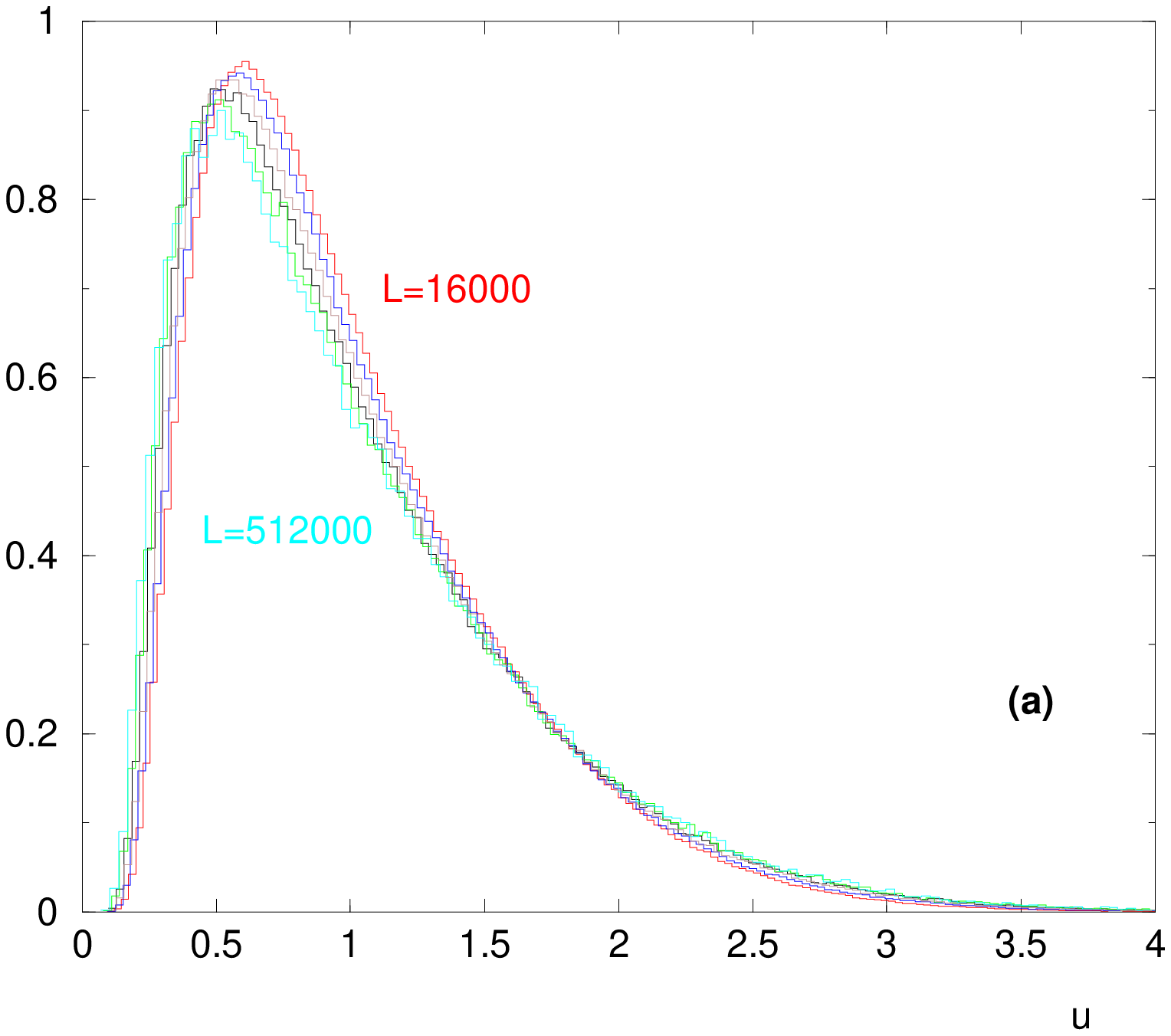}
\hspace{1cm}
\includegraphics[height=6cm]{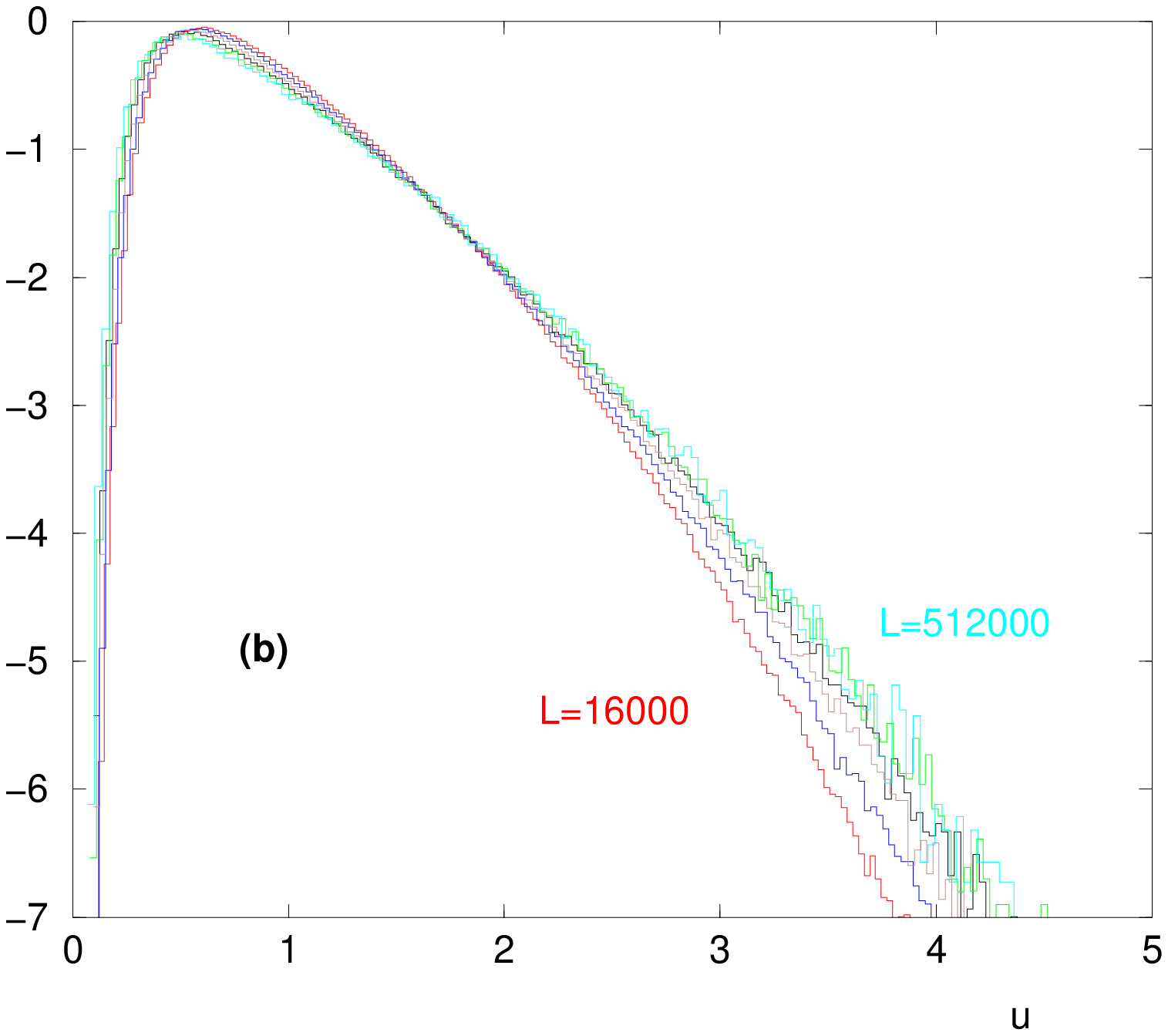}
\caption{(Color online) Wetting with loop exponent $c=1.75$ :
Wiseman-Domany lack of self-averaging at criticality
for the spatial average $\rho(i,L)$ of the order parameter (Eq. \ref{rhospatialav})
(a) histogram over the samples of the variable
$u=\rho(i,L)/\overline{\rho(i,L) }$ for sizes $16.10^3 \leq L \leq
512.10^3$ :  $u$ remains a random variable of order $O(1)$ 
in the limit $L \to \infty$.
(b) same data in logarithmic scale. }
\label{fighistodomanyc1.75}
\end{figure}

To study the Wiseman-Domany lack of self-averaging at criticality
for the spatial average $\rho(i,L)$ of order parameter (Eq. \ref{rhoaviq}),
we have computed the probability distribution $G_L(u)$ of
the ratio 
\begin{equation}
u = \frac{\rho(i,L)}{\overline{\rho(i,L) }} 
\label{udomanyc1.75}
\end{equation}
The results for various $L$ presented on Fig \ref{fighistodomanyc1.75}
show that $u$ remains a random variable of order $O(1)$ 
in the limit $L \to \infty$.

\section{Multifractal analysis of the wetting transition 
with loop exponent $c=1.5$ }

 \label{resc1.5}

In this Section, we describe our results for the wetting transition
with loop exponent $c=1.5$ that corresponds to marginal disorder
as explained above
(Eq. \ref{harriscriterion}).

\subsection{Exponents $x(q)$ and generalized dimensions $D(q)$ 
and $ \tilde D(q)$ }

\begin{figure}[htbp]
\includegraphics[height=6cm]{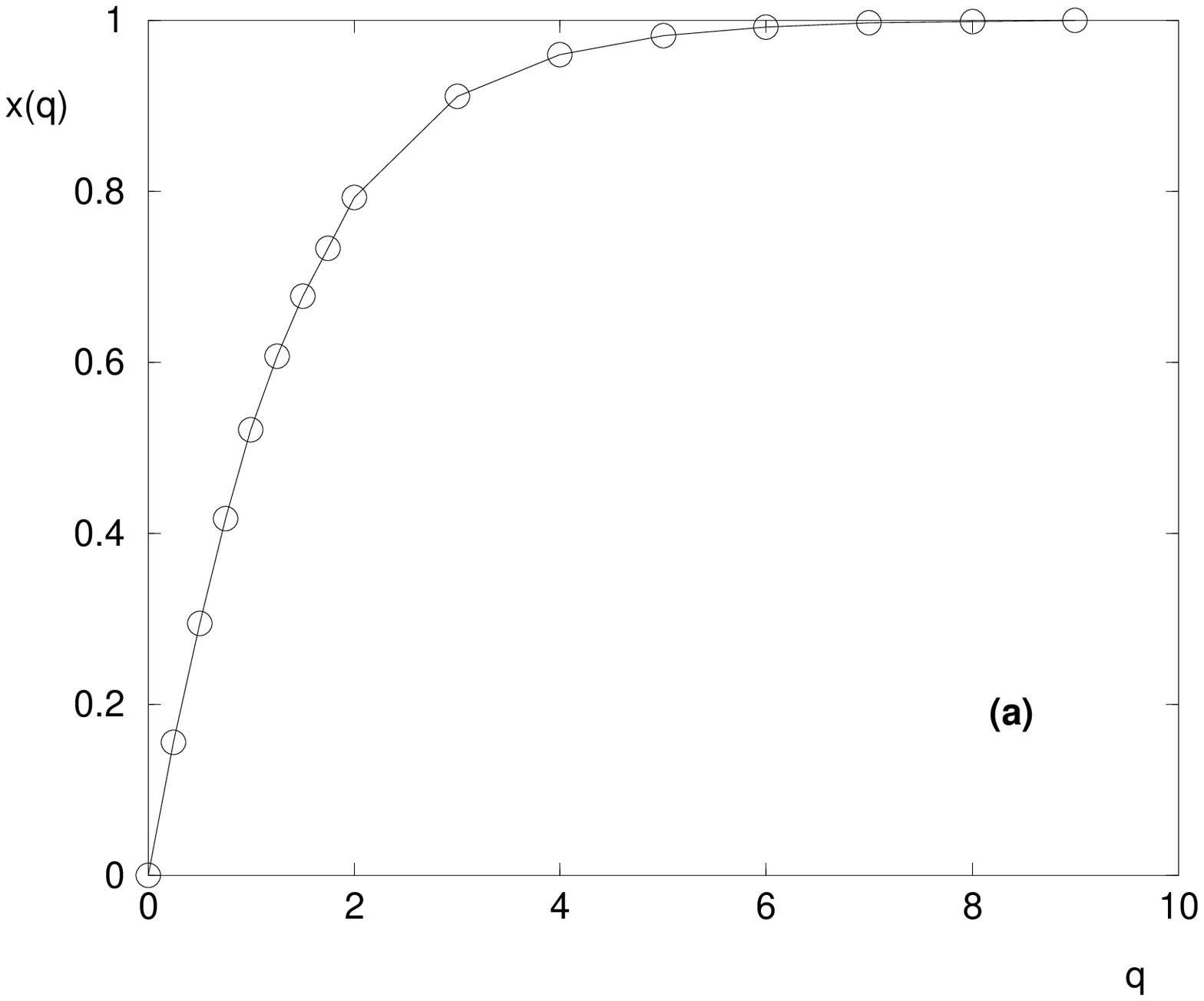}
\hspace{1cm}
\includegraphics[height=6cm]{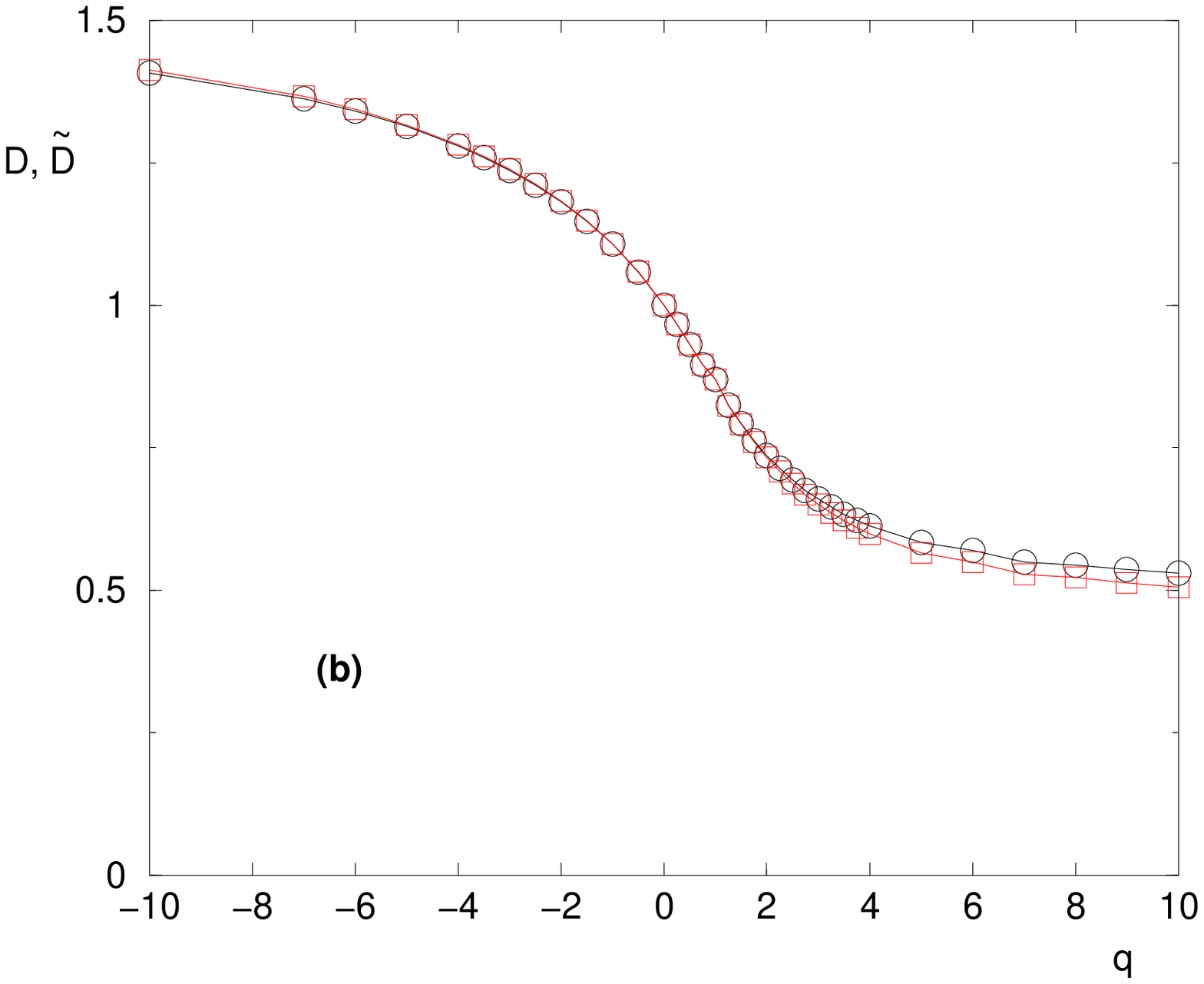}
\caption{(Color online) Wetting with loop exponent $c=1.5$
(a) Exponents $x(q)$ governing the decay of the disorder averaged $q$-th powers of the
local order parameter (Eq (\ref{defxqwettbis})
(b) Generalized dimensions $D(q)$ ($\bigcirc$) and $\tilde D(q)$ ($\square$) associated respectively
to the typical values (Eq. \ref{tctyp}) and 
to disorder-averaged values (Eq. \ref{tcav}) of the generalized moments
(Eq. \ref{yqilwettbis}). }
\label{figexpc1.5}
\end{figure}

We show on Fig. \ref{figexpc1.5} (a) 
the scaling dimensions $x(q)$ governing
the disorder-averaged moments of the local contact density
(Eq. \ref{rhoqav})
\begin{eqnarray}
 \overline{ < \delta_{z_r,0} >^q } \propto \frac{1}{ L^{x(q)} }
\label{defxqwettbis}
\end{eqnarray}
It is strongly non linear $x(q) \ne q x(1)$.
Some particular values are
\begin{eqnarray}
x(q=1) && \simeq 0.52 \\
x(q=2) && \simeq 0.79
\end{eqnarray}
In the large $q$ limit, it saturates towards
\begin{eqnarray}
x(q \to +\infty) \to 1 
\label{xqinfty2}
\end{eqnarray}
(see the discussion of Section \ref{boundary}).

On Fig. \ref{figexpc1.5} (b) we show the generalized dimensions
$D(q)$ and $\tilde D(q)$ associated respectively
to the typical values (Eq. \ref{tctyp}) and 
to disorder-averaged values (Eq. \ref{tcav}) of the generalized moments
(Eq. \ref{defyq})
\begin{eqnarray}
Y_q(i,L) = \frac{ \displaystyle  \sum_{r} < \delta_{z_r,0} >^q }
{ \left( \displaystyle  \sum_{r' } < \delta_{z_r',0} >  \right)^q}  
\label{yqilwettbis}
\end{eqnarray}
In particular, the 'information dimension of Eq. \ref{entropy}
is
\begin{eqnarray}
D(q=1) = {\tilde D}(1) \simeq 0.87
\end{eqnarray}
and the correlation dimension 
of Eq. \ref{y2d2} is
\begin{eqnarray}
D(q=2) \sim {\tilde D}(2) \simeq 0.74
\end{eqnarray}

\subsection{Typical singularity spectrum $f(\alpha)$ }

\begin{figure}[htbp]
\includegraphics[height=6cm]{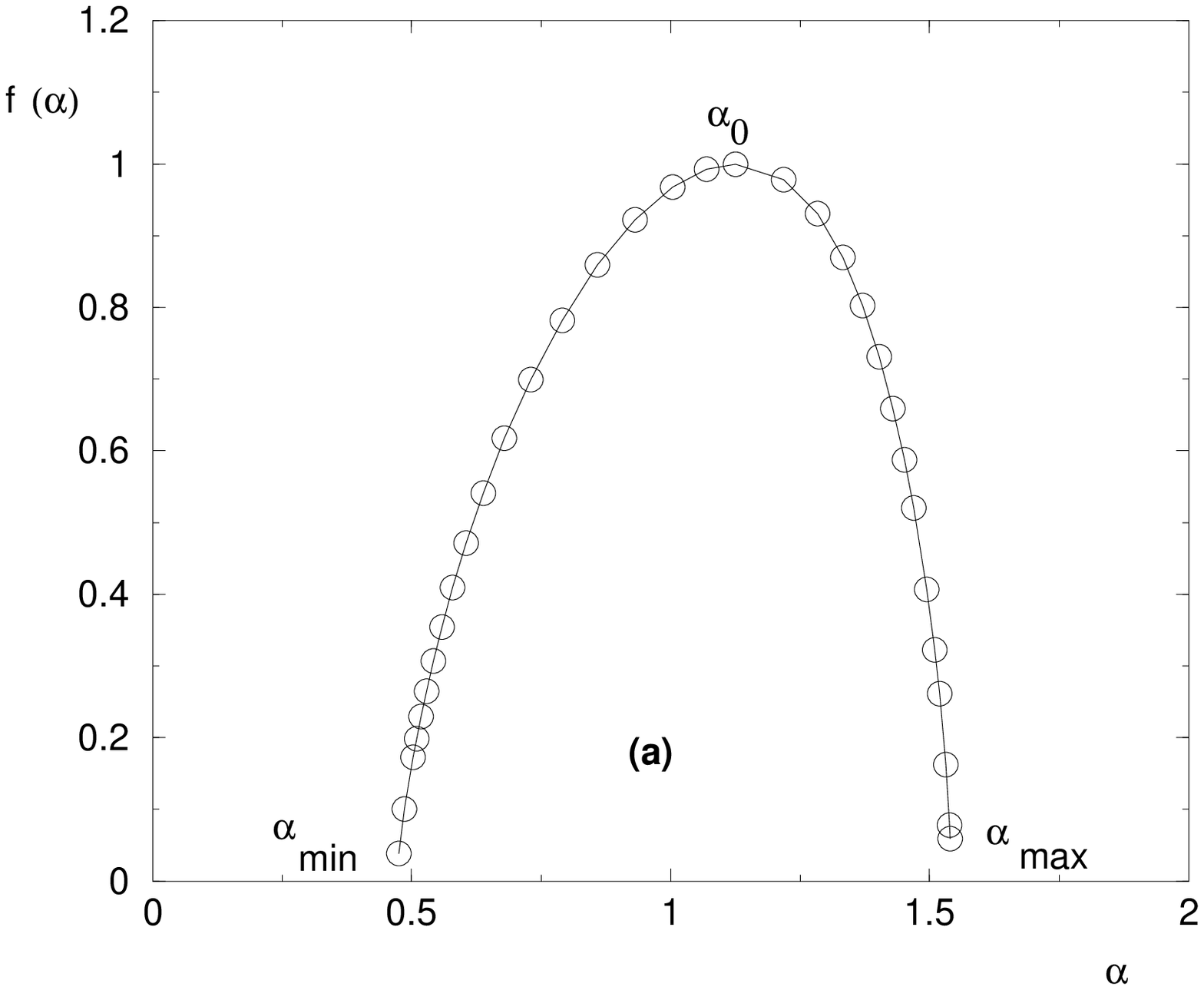}
\hspace{1cm}
\includegraphics[height=6cm]{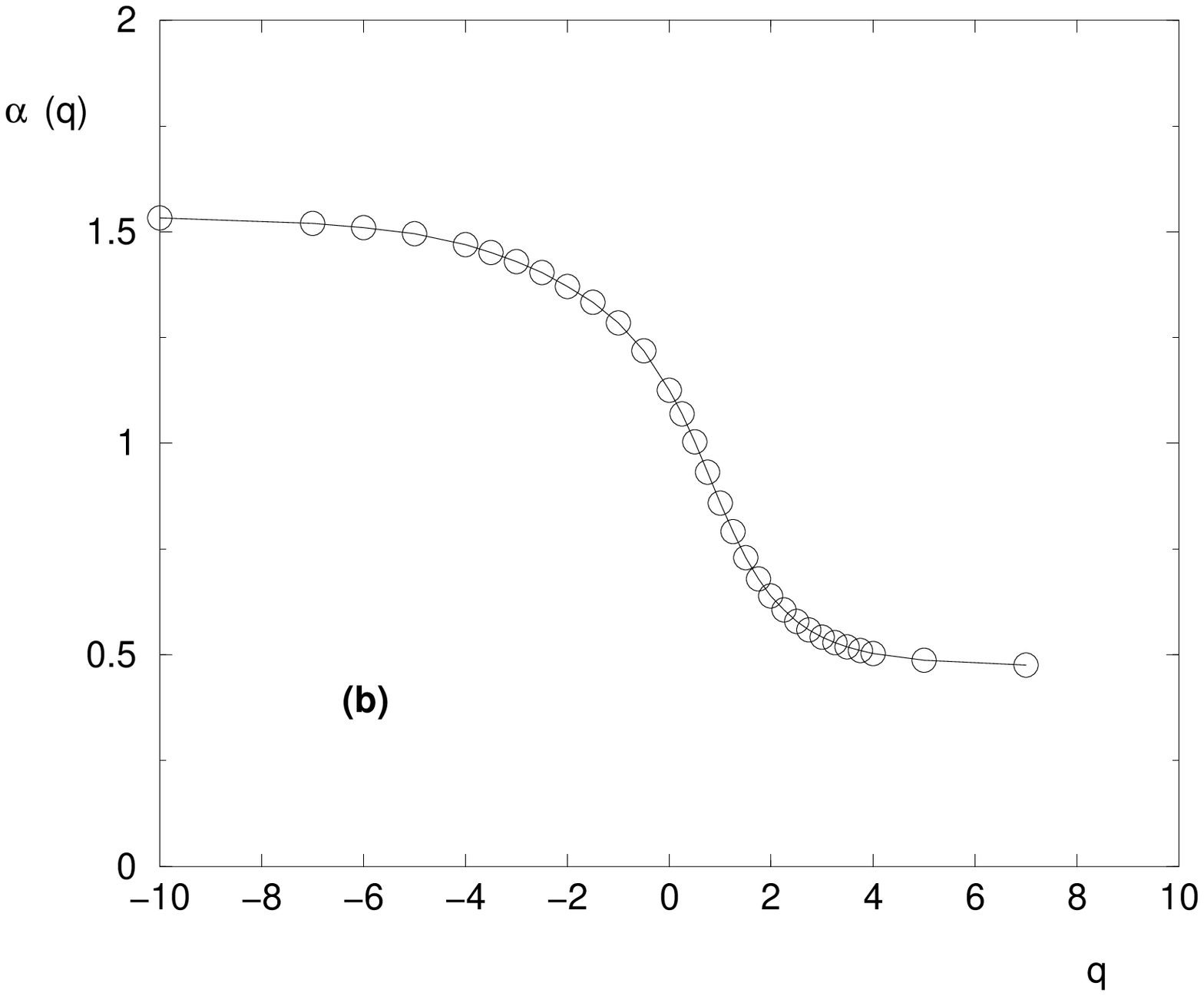}
\caption{(Color online) Wetting with loop exponent $c=1.5$
(a) Typical singularity spectrum $f(\alpha)$ (Eq. \ref{nlalpha}) :
the maximum occurs at $\alpha_0 \sim 1.125 $ which represents the
typical value.
The minimal value is around $\alpha_{min} \simeq 0.48$.
(b) Corresponding curve $\alpha(q)$.  }
\label{figfalphac1.5}
\end{figure}

We show on Fig. \ref{figfalphac1.5} (a) 
the curve $f(\alpha)$ obtained via the standard method 
 of Ref. \cite{Chh}
The maximum corresponds to the typical exponent $\alpha_0$ (Eq.
\ref{alphatyp})
\begin{equation}
\alpha_{typ}=\alpha_0 = 1.125
\label{alphatypc1.5}
\end{equation}
The curve is tangent to the diagonal $\alpha=f(\alpha)$ at the point (Eq. \ref{alpha1})
\begin{equation}
\alpha_1=f(\alpha_1)= D(1) \simeq 0.87
\label{alpha1c1.5}
\end{equation}
The minimal value corresponds to
\begin{equation}
\alpha_{min}= D(q=+\infty) \simeq 0.48
\end{equation}
and the maximal value
\begin{equation}
\alpha_{max}= D(q=-\infty) \simeq 1.53
\end{equation}

On Fig. \ref{figfalphac1.5} (b), we show the corresponding curve $\alpha(q)$
representing the dominant exponent $\alpha$ that contribute
to the $q$-generalized moment (Eq. \ref{saddle}).

\subsection{ Disorder-averaged singularity spectrum ${\tilde f}(\alpha)$  }

\begin{figure}[htbp]
\includegraphics[height=6cm]{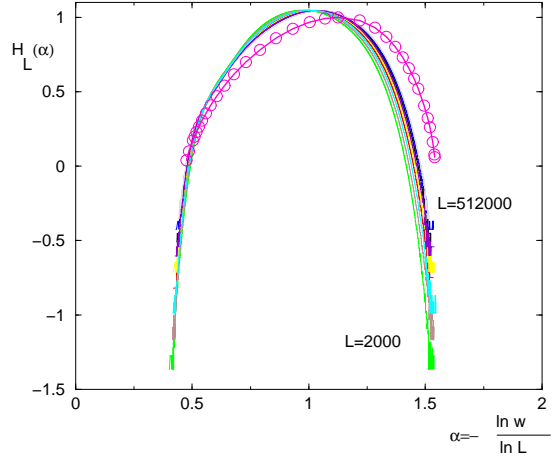}
\hspace{1cm}
\caption{(Color online) Wetting with loop exponent $c=1.5$ :
the histograms $H_L(\alpha)$ of the rescaled variable $\alpha = -
\frac{ \ln w(r;i,L) }{ \ln L}$
shown for $2.10^3 \leq L \leq 512.10^3$
converge extremely slowly (as $1/\ln(L)$) towards the typical singularity spectrum
$f(\alpha)$ ($\bigcirc$) measured with the method of Ref. \cite{Chh}.
 }
\label{figftildec1.5}
\end{figure}

We show on Fig. \ref{figftildec1.5} the histograms $H_L(\alpha)$ of 
\begin{equation}
\alpha = - \frac{ \ln w(r;i,L) }{ \ln L}
\end{equation}
for various sizes $L$ and compare with the 
the typical spectrum $f(\alpha)$
obtained via the method of Ref. \cite{Chh}.
As previously mentioned for the corresponding Figure \ref{figftildec1.75}
concerning the case $c=1.75$, the convergence is extremely
slow, of order $1/\ln(L)$.

\subsection{ Histograms of the `information' entropy and of $Y_2$  }

\begin{figure}[htbp]
\includegraphics[height=6cm]{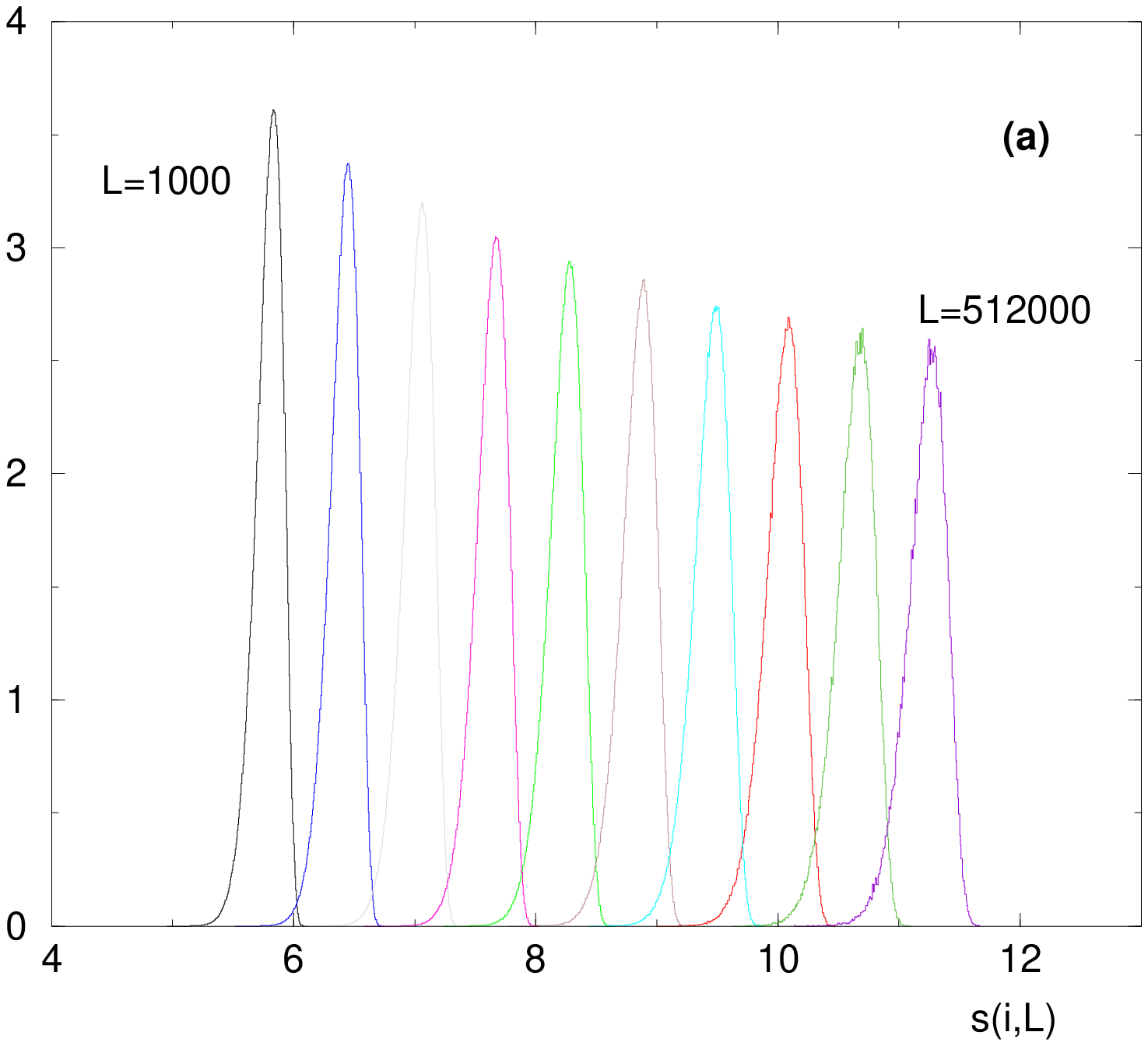}
\hspace{1cm}
\includegraphics[height=6cm]{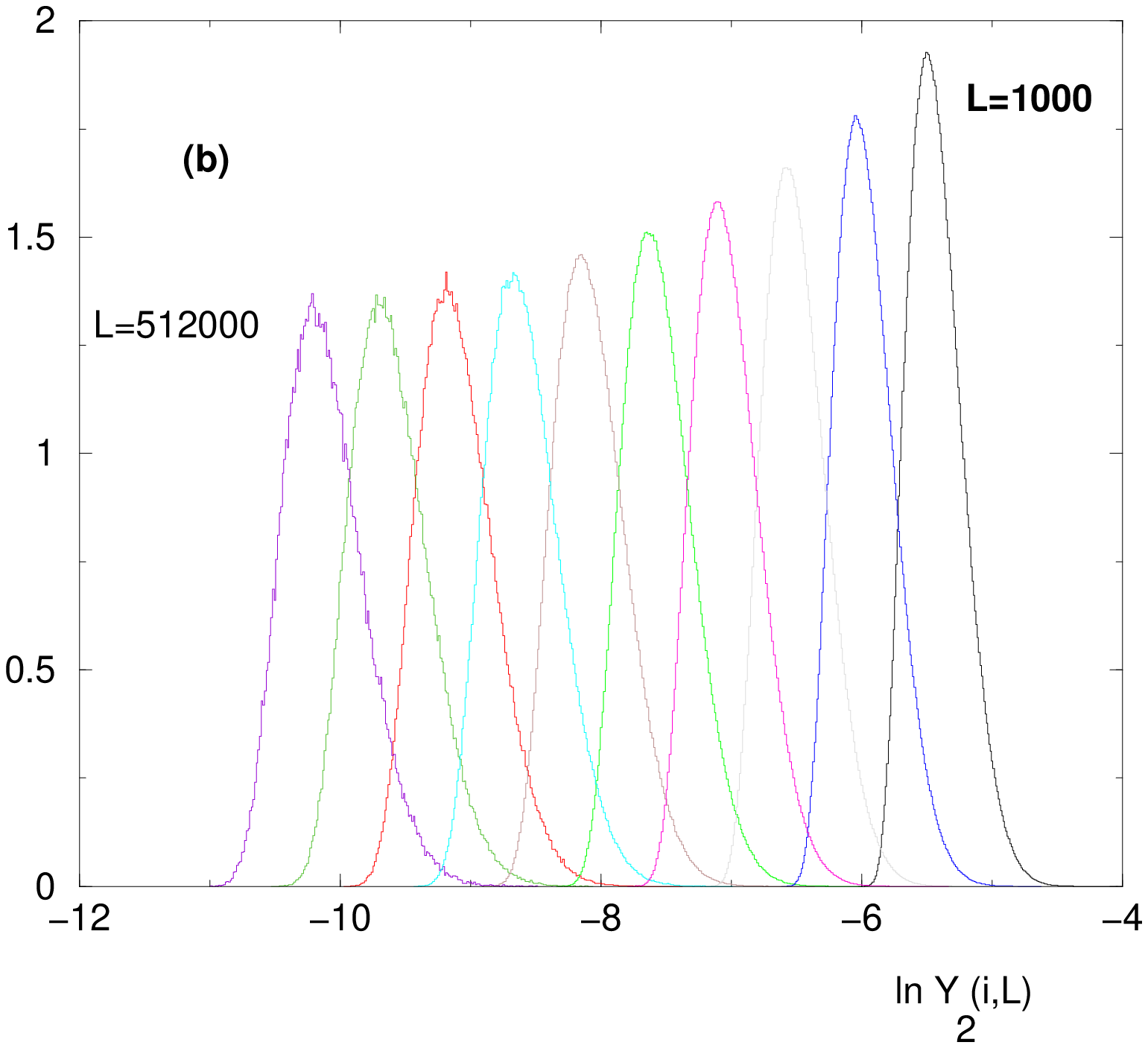}
\caption{(Color online) Wetting with loop exponent $c=1.5$
(a) histogram over the samples of the information entropy
$s(i,L)$ defined in Eq. \ref{entropy} for sizes $10^3 \leq L \leq
512.10^3$ :  the averaged value grows as $\overline{s(i,L) }
\simeq  D(1) \ln L$
(b) histogram over the samples of $\ln Y_2(i,L)$ (Eq.  \ref{defyq} )
 for sizes $10^3 \leq L \leq
512.10^3$ :  the averaged value behaves as $\overline{  \ln Y_2(i,L) }
\simeq  - D(2) \ln L$. } 
\label{fighistoc1.5}
\end{figure}

We show on Fig. \ref{fighistoc1.5} (a) the histogram over the samples $(i)$
of the 'information' entropy $s(i,L)$ defined in Eq. \ref{entropy} :
as $L$ grows, the averaged value grows logarithmically (Eq.  \ref{entropy})
\begin{equation}
\overline{s(i,L) } \simeq  D(1) \ln L
\label{entropyav2}
\end{equation}
whereas the width again converges towards a constant value.

Similarly, we show on Fig. \ref{fighistoc1.5} (b) 
the histogram over the samples $(i)$
of $\ln Y_2(i,L)$ (Eq.  \ref{defyq} ) :
as $L$ grows, the averaged value grows logarithmically (Eq. \ref{y2d2})
\begin{equation}
\overline{  \ln Y_2(i,L) } \simeq  - D(2) \ln L
\label{lny2av2}
\end{equation}
whereas the width converges towards a constant value.

\subsection{Statistics of the spatial averaged order parameter
 over the samples }

\begin{figure}[htbp]
\includegraphics[height=6cm]{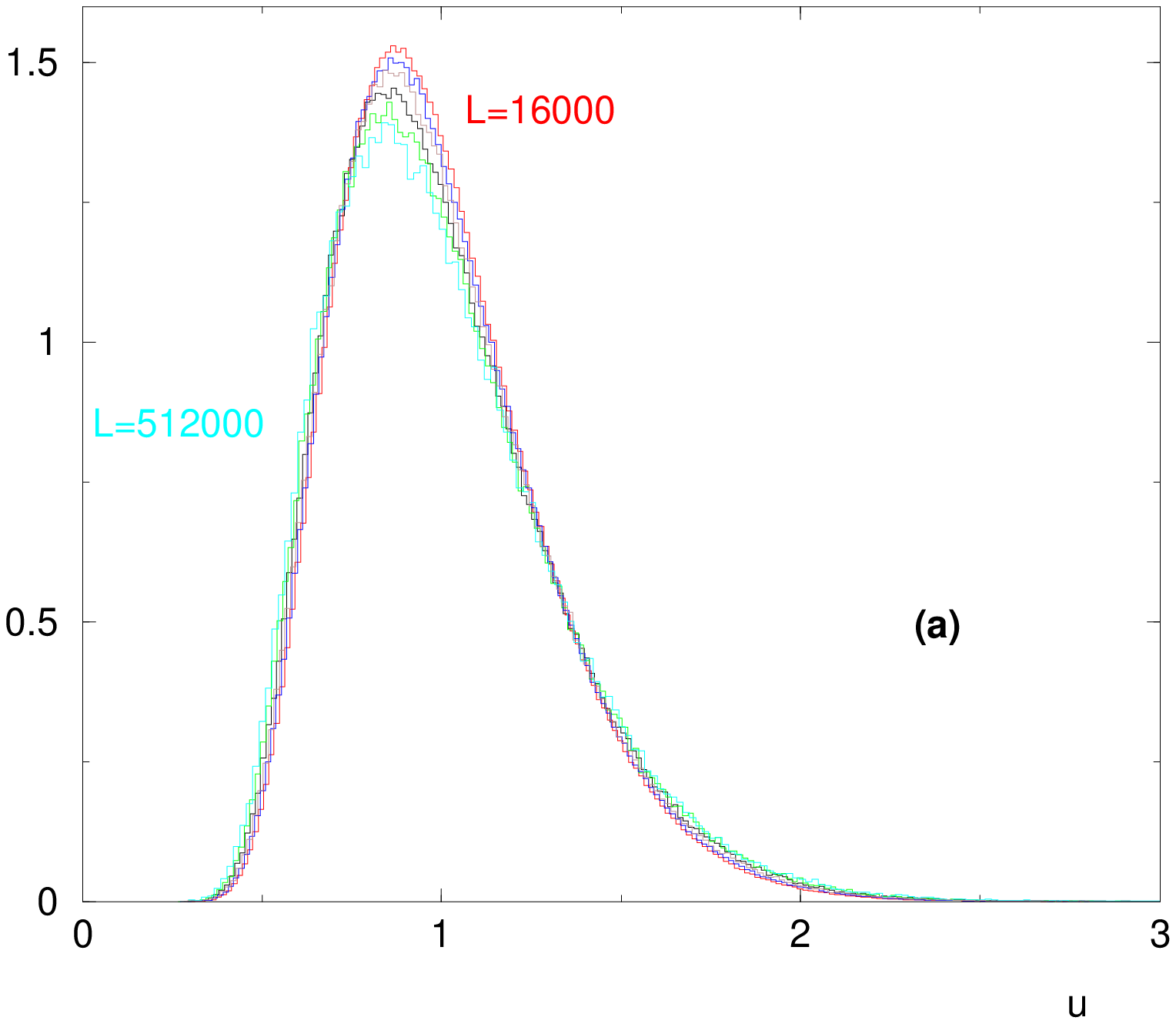}
\hspace{1cm}
\includegraphics[height=6cm]{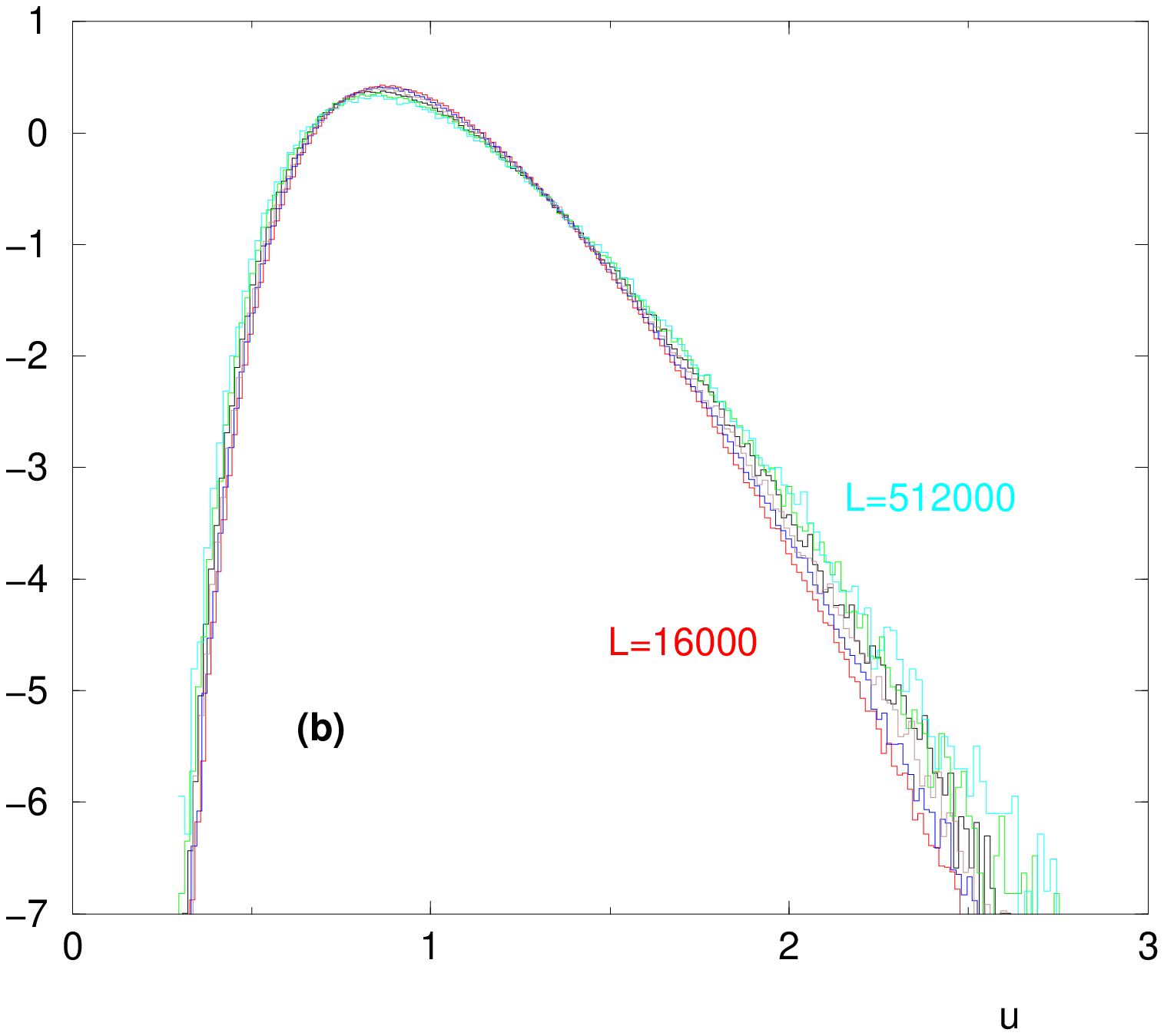}
\caption{(Color online) Wetting with loop exponent $c=1.5$ :
Wiseman-Domany lack of self-averaging at criticality  
for the spatial average $\rho(i,L)$ of the order parameter
 (Eq. \ref{rhospatialav})
(a) histogram over the samples of the variable
$u=\rho(i,L)/\overline{\rho(i,L) }$ for sizes $16.10^3 \leq L \leq
512.10^3$ :  $u$ remains a random variable of order $O(1)$ 
in the limit $L \to \infty$.
(b)  same data in logarithmic scale.  }
\label{fighistodomanyc1.5}
\end{figure}

To study the Wiseman-Domany lack of self-averaging at criticality
for the spatial average $\rho(i,L)$ of order parameter (Eq. \ref{rhoaviq}),
we have computed the probability distribution $G_L(u)$ of
the ratio 
\begin{equation}
u = \frac{\rho(i,L)}{\overline{\rho(i,L) }} 
\label{udomanyc1.5}
\end{equation}
The results for various $L$ are shown on Fig \ref{fighistodomanyc1.5}.
show that $u$ remains a random variable of order $O(1)$ 
in the limit $L \to \infty$.

\section{ Influence of boundary conditions }

 \label{boundary}

\subsection{ Value $x(q \to \infty)$ and minimal value $\alpha_{min}$
 of the multifractal spectrum }

In our numerical studies for $c=1.75$ and $c=1.5$ presented above,
we have found in both cases 
that the exponents $x(q)$ governing the decay of the powers of the
local order parameter (Eq \ref{rhoqav})
saturate to $1$ for large $q$ (Eqs \ref{xqinfty1} and \ref{xqinfty2})
\begin{eqnarray}
x(q \to \infty) =1
\label{xqinfty}
\end{eqnarray}
This indicate that moments of high order $q \to \infty$
are dominated by a finite number of points (of density of order $1/L$)
having a finite order parameter $O(1)$.
So the minimal $y_{min}$ actually saturates
the bound of Eq. \ref{ymin}
\begin{eqnarray}
y_{min}= \alpha_{min}-d+x(1) = 0
\label{yminzero}
\end{eqnarray}
As a consequence, the minimal value $\alpha_{min}$ of 
the multifractal spectrum is simply given by
\begin{eqnarray}
 \alpha_{min}=d-x(1) 
\end{eqnarray}
This simple relation is satisfied with the above numerical results
both for $c=1.75$ (with $x(q=1) \simeq 0.63$ and 
$\alpha_{min}= D(q=+\infty) \simeq 0.36$)
and for $c=1.5$ ( with $x(q=1) \simeq 0.52$ and 
$\alpha_{min}= D(q=+\infty) \simeq 0.48$ )

To determine whether these properties are linked to
the bound-bound boundary conditions used, or are more general,
we have studied other boundary conditions as we now explain.

\subsection{ Comparison between fixed and free boundary conditions  }

\begin{figure}[htbp]
\includegraphics[height=6cm]{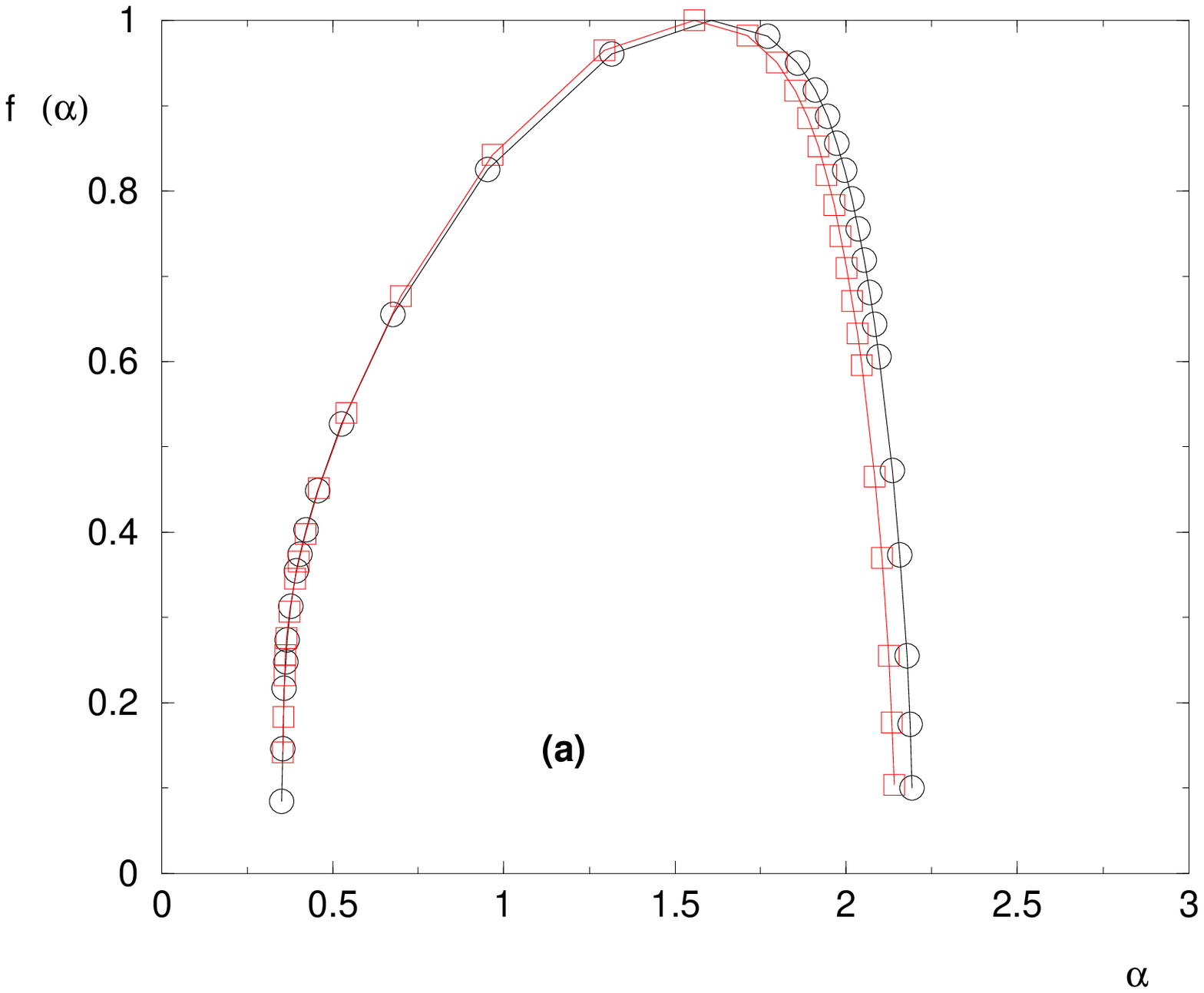}
\hspace{1cm}
\includegraphics[height=6cm]{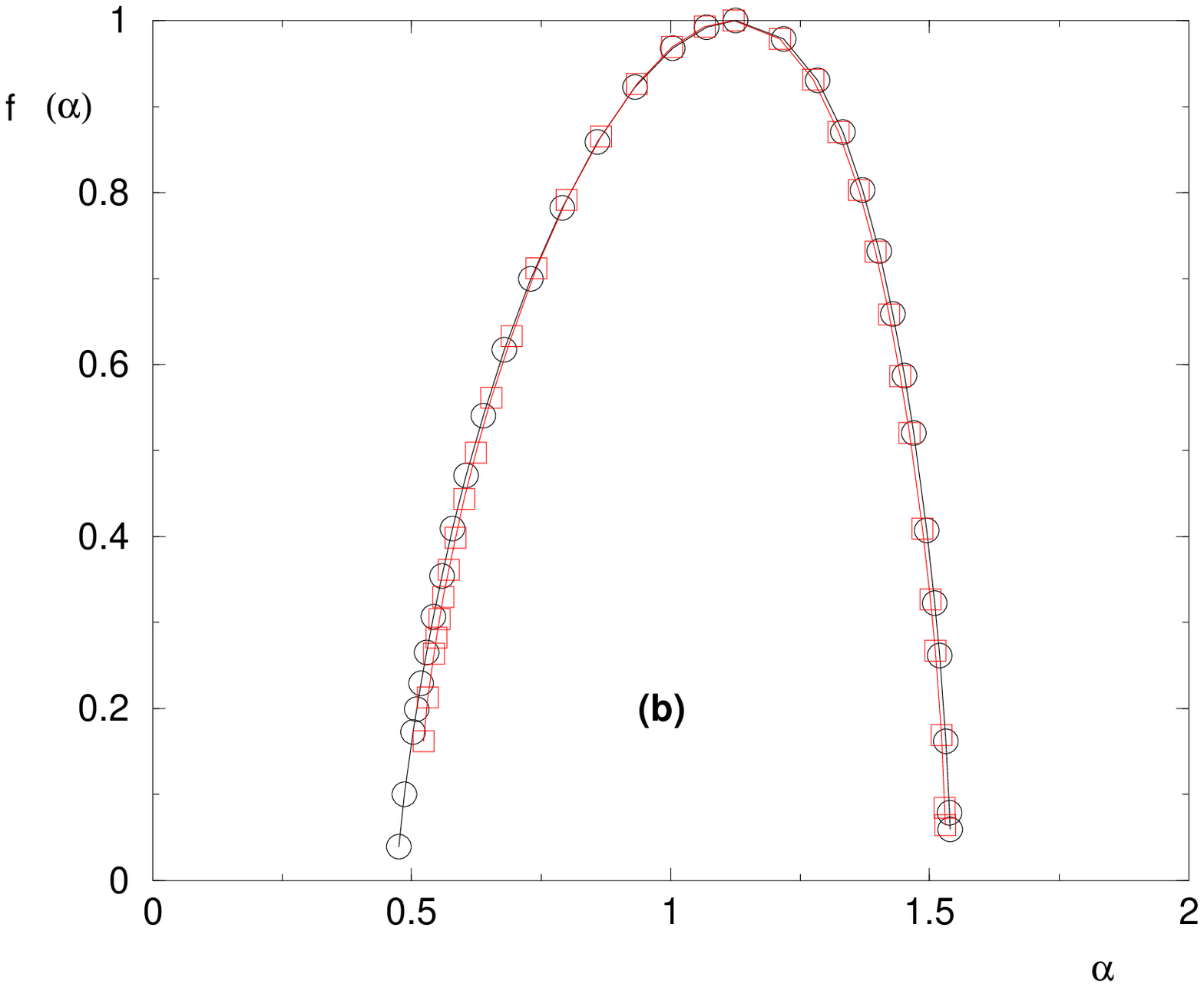}
\caption{(Color online) Typical singularity spectra $f(\alpha)$ for bound-bound 
($\bigcirc$) or free-free ($\square$)
boundary conditions
(a) wetting with loop exponent $c=1.75$ (b) wetting with loop exponent
$c=1.5$. } 
\label{figfalphabulk}
\end{figure}

In the previous Sections \ref{resc1.75} and \ref{resc1.5},
we have presented numerical results concerning the bound-bound boundary conditions,
where the polymer is attached to the interface $z=0$ at both ends
\begin{eqnarray}
 z(r=1)=0=z(r=L)
\label{boundbound}
\end{eqnarray}
Since the wetting models cannot be directly defined with
completely free boundary conditions (since the space above the interface is infinite),
we have still considered the bound-bound boundary conditions of
Eq. \ref{boundbound} but we have measured the multifractal spectrum
using only the `bulk monomers' satisfying 
\begin{eqnarray}
\frac{L}{4} \leq r \leq \frac{3 L}{4}
\label{rbulk}
\end{eqnarray}
This procedures aims to simulate 'free' boundary conditions
at $r=\frac{L}{4}$ and at $r=\frac{3 L}{4}$ for a sample of size 
 $\frac{L}{2}$.
The singularity spectra measured with these two types of boundary
conditions are shown on Fig. \ref{figfalphabulk} (a) and (b) for
the cases $c=1.75$ and $c=1.5$ respectively.
These singularity spectra coincide within our numerical accuracy
(the difference on the right half of Fig. \ref{figfalphabulk} (a)
corresponds to the negative moments $q<0$ dominated by the smallest weights
whose statistics is more difficult to measure precisely).

\subsection{ Griffiths ordered clusters  }

In particular, the fact that the minimal value $\alpha_{min}$
remains the same for fixed and 'free' boundary conditions
shows that the finite number of points having
a finite order parameter are not confined to the boundaries
but also exist in the bulk at criticality.
This is related to the finite probability of finite ordered clusters
in diluted disordered systems below the pure critical temperature,
which have been much studied in the context of Griffiths singularities
\cite{Gri}.

Our conclusion for the wetting transition and more generally for disordered systems at
criticality, is that the finite probability of Griffiths ordered clusters
determines the asymptotic value
\begin{eqnarray}
x(q \to \infty) =d
\label{xqinftygen}
\end{eqnarray}
of the exponents $x(q)$ governing the decay of the powers of the
local order parameter.
Accordingly, the minimal $y_{min}$ saturates
the bound of Eq. \ref{ymin}
\begin{eqnarray}
y_{min}= \alpha_{min}-d+x(1) = 0
\label{yminzerogene}
\end{eqnarray}
and the minimal value $\alpha_{min}$ of 
the multifractal spectrum is simply related to the exponent $x(1)$
governing the decay of the order parameter
\begin{eqnarray}
 \alpha_{min}=d-x(1) 
\end{eqnarray}

\section{ Summary and conclusions } 

 \label{conclusion}

In this paper, we have studied in detail the 
multifractal statistics of the local order parameter 
at random wetting critical points, for two values of the loop exponents $c$,
namely $c=1.75$ and $c=1.5$.
For these models where large sizes up to $L=512.10^3$ can be probed with
a good statistics over the samples, we have numerically measured \\
(i) the 'Ludwig exponents' $x(q)$
that govern the moments $\overline{\rho^q(r)}$ 
of the local order parameter $\rho(r)$ \\
(ii) the generalized dimensions $D(q)$ and $\tilde D(q)$ associated
to typical and disorder-averaged values of the moments $Y_q =\sum_r w^q(r)$
of the multifractal measure $w(r) = \rho(r)/(\sum_{r'} \rho(r'))$. \\
(iii) the corresponding typical singularity spectrum $f(\alpha)$.

We have also discussed the relations between this multifractal
statistics and the Wiseman-Domany lack of self-averaging at criticality.
Finally, we have argued that the presence of finite Griffiths
ordered clusters at $T_c$ determines the asymptotic value
of the Ludwig exponent $x(q \to \infty) =d$
and the minimal value $\alpha_{min}=D(q \to \infty)=d-x(1) $
of the multifractal spectrum.
We have checked that these relations are well satisfied 
in our numerical results for the wetting transitions (where $d=1$), 
not only for bound-bound boundary conditions
but also for 'free-free' boundary conditions.

However finite Griffiths ordered clusters occur above 
critical points in disordered systems independently of
 the relevance/irrelevance of the disorder from the Harris criterion
\cite{Gri}, which concerns coarse-grained properties at weak disorder. 
Our conclusion is thus that the multifractal statistics
of the local order parameter should be non-trivial for any critical
point with frozen disorder, since it probes the heterogeneities at all
scales. This would explain why our results concerning the marginal
disorder case $c=1.5$  are qualitatively similar to our results for
the relevant disorder case $c=1.75$.


\begin{thebibliography}{99}





\bibitem{halsey}
T.C. Halsey, M.H. Jensen, L.P. Kadanoff, I. Procaria and B. Shraiman,
Phys. Rev. A 33, 1141 (1986).

\bibitem{Pal_Vul}
G. Paladin and A. Vulpiani, Phys. Rep. 156, 147 (1987).

\bibitem{Stan_Mea}
H.E. Stanley and P. Meakin, Nature 335, 405 (1988).

\bibitem{Aha}
A. Aharony and J. Feder Eds, {\it Fractals in Physics}, Essays in
honour of B.B. Mandelbrot, North Holland (1990).

\bibitem{Meakin}
P. Meakin, {\it Fractals, scaling and growth far from equilibrium},
Cambridge (1998).


\bibitem{harte}
D. Harte, '' Multifractals, Theory and Applications'', Chapman and Hall (2001).


\bibitem{duplantier_houches}
 B. Duplantier, ''Conformal Random Geometry'',
 Les Houches, Session LXXXIII, 2005, Mathematical Statistical Physics,
Eds A. Bovier et al., 101, Elsevier  (2006).


\bibitem{Weg}
F. Wegner, Z. Phys. B 36, 209 (1980).

\bibitem{Cas_Pel}
C. Castellani and L. Peliti, J. Phys. A 19, L429 (1986)

\bibitem{Jan}
M. Janssen, Int. J. Mod. Phys. 8, 943 (1994);
M. Janssen, Phys. Rep. 295, 1 (1998).

\bibitem{Huck}
B. Huckestein, Rev. Mod. Phys. 67, 357 (1995).

\bibitem{Mirlin}
F. Evers and A.D. Mirlin, Phys. Rev. Lett. 84 , 3690 (2000) ;
 A.D. Mirlin and F. Evers, Phys. Rev. B 62, 7920 (2000);
F. Evers, A. Mildenberger and A.D. Mirlin, Phys. Rev. B 64, 241003
(2001);
 A. Mildenberger, F. Evers, and A. D. Mirlin
Phys. Rev. B 66, 033109 (2002);
A. D. Mirlin, Y. V. Fyodorov, A. Mildenberger, and F. Evers
Phys. Rev. Lett. 97, 046803 (2006).


\bibitem{DPmultif}
C. Monthus and T. Garel, cond-mat/0701699.


\bibitem{Ludwig}
A.W.W. Ludwig, Nucl. Phys. B 330, 639 (1990).




\bibitem{Jac_Car}
J.L. Jacobsen and J.L. Cardy, Nucl. Phys., B515, 701 (1998).

\bibitem{Ols_You}
T. Olsson and A.P. Young, Phys. Rev., B60, 3428 (1999).

\bibitem{Cha_Ber}
C. Chatelain and B. Berche, Nucl. Phys., B572, 626 (2000).

\bibitem{PCBI}
G. Pal\'agyi, C. Chatelain, B, Berche and F. Igl\'oi, Eur. Phys. J,
B13, 357 (2000).




\bibitem{Sourlas}
N. Sourlas, Europhys. Lett. 3, 1007 (1987).

\bibitem{Thi_Hil}
M.J. Thill and H.J. Hilhorst, J. Phys. I 6, 67 (1996).



\bibitem{Par_Sou}
G. Parisi and N. Sourlas, Phys. Rev. Lett. 89, 257204 (2002).

\bibitem{brazil}
E. Nogueira Jr., S. Coutinho, F.D. Nobre, E.M.F. Curado and J.R.L. de
Almeida, Phys. Rev., E55, 3934 (1997).

\bibitem{multiscaling}
J. Kisker and A. P. Young
Phys. Rev. B 58, 14397-14400 (1998) ;
F. Igloi, R. Juhasz, and H. Rieger
Phys. Rev. B 61 11552 (2000).

\bibitem{revueigloi}
F. Igl\'oi and C. Monthus, Phys. Rep. 412, 277 (2005).


\bibitem{domany95}
S. Wiseman and E. Domany,
Phys Rev E {\bf 52}, 3469 (1995).

\bibitem{AH}
A. Aharony, A.B. Harris,
Phys Rev Lett {\bf 77}, 3700 (1996).

\bibitem{domany}
S. Wiseman and E. Domany,
Phys. Rev. Lett. {\bf 81}, 22 (1998) ; Phys Rev E {\bf 58}, 2938 (1998).

\bibitem{Dup_Lud}
B. Duplantier and A.W.W. Ludwig, Phys. Rev. Lett. 66, 247 (1991).

\bibitem{Pook}
W. Pook and M. Janssen, Z. Phys. B 82, 295 (1991).




\bibitem{mandelbrot}
B. Mandelbrot, Physica A 163, 306 (1990)
B. Mandelbrot, J. Stat. Phys. 110, 739 (2003)

\bibitem{Chh_neg}
A.B. Chhabra and K.R. Sreenivasan, Phys. Rev. A 43, 1114 (1991).


\bibitem{Jen_neg}
M.H. Jensen, G. Paladin and A. Vulpiani, Phys. Rev. E 50 , 4352 (1994).

\bibitem{has_dup}
T.C. Halsey, K. Honda and B. Duplantier, J. Stat. Phys. 85, 681 (1996);
T.C. Halsey, B. Duplantier and K. Honda, Phys. Rev. Lett. 78, 1719 (1997). 

\bibitem{Harris}
A.B. Harris, J. Phys. C {\bf  7}, 1671 (1974).




\bibitem{mfisher}
M. E. Fisher, J. Stat. Phys. 34 (1984) 667 .

\bibitem{Pol_Scher} D. Poland and H.A. Scheraga eds., Academic Press, New York (1970) 
``Theory of Helix-Coil transition in Biopolymers''.

\bibitem{Fisher}
M. Fisher, J. Chem. Phys. 45 (1966) 1469 .

 \bibitem{Barbara1}
M.S. Causo, B. Coluzzi and P. Grassberger
Phys. Rev. E {\bf 62}, 3958 (2000) .



\bibitem{Carlon} 
E. Carlon, E. Orlandini and
A.L. Stella, Phys. Rev. Lett., {\bf 88}, 198101 (2002) ;
M. Baiesi, E. Carlon, and A.L. Stella, Phys. Rev. E,
{\bf 66}, 021804 (2002) ;
M. Baiesi, E. Carlon, Y. Kafri, D. Mukamel,
E. Orlandini and A.L. Stella, Phys. Rev. E,
{\bf 67}, 021911 (2002) .


\bibitem{Ka_Mu_Pe} Y. Kafri, D. Mukamel and L. Peliti, Phys. Rev. Lett., 
\textbf{85}, 4988 (2000) : Y. Kafri, D. Mukamel and L. Peliti, Eur. Phys. J. B,
{\bf 27}, 135 (2002).


\bibitem{harris}
A. B. Harris,
J. Phys. C  7 , 1671 ( 1974).


\bibitem{FLNO} 
G. Forgacs, J.M. Luck, Th.M. Nieuwenhuizen and
H. Orland, Phys. Rev. Lett., {\bf 57}, 2184 (1986) 
and  J. Stat. Phys., {\bf 51}, 29 (1988).

\bibitem{Der_Hak_Van}
B. Derrida, V. Hakim, J. Vannimenus,  J. Stat. Phys. 66, 1189 (1992)

\bibitem{Bhat_Muk} 
S.M. Bhattacharjee and S. Mukherjee, Phys. Rev. Lett., 
\textbf{70}, 49 (1993); Phys. Rev. {\bf E48}, 3483 (1993).


\bibitem{Ka_La} 
H. Kallabis and M. L\"assig, Phys. Rev. Lett., 
\textbf{75}, 1578 (1995)

\bibitem{Cu_Hwa} 
D. Cule and T. Hwa, Phys. Rev. Lett., 
\textbf{79}, 2375 (1997)

\bibitem{Ta_Cha}
       L.-H. Tang and H. Chat\'e
       Phys. Rev. Lett. 86, 830 (2001).

\bibitem{wetting2005}
T. Garel and C. Monthus, Eur. Phys. J. B 46, 117 (2005).

 \bibitem{PS2005}
  C.Monthus and T. Garel, Eur. Phys. J. B 48, 393-403 (2005).

\bibitem{Gia}
G. Giacomin
``Random Polymer Models''
Imperial College Press (2006).


\bibitem{Chh}
A. Chhabra and R.V. Jensen, Phys. Rev. Lett. 62, 1327 (1989).



\bibitem{Gri}
R.B. Griffiths, Phys. Rev. Lett. 23, 17 (1969);
A.B. Harris, Phys. Rev. B 12, 203 (1975);
A.J. Bray, Phys. Rev. Lett. 59, 586 (1987);
T. Vojta, J.Phys. A 39, R143 (2006).


\end{thebibliography}
\end{document}